\documentclass[11pt,preprint]{aastex}
\slugcomment{submitted to ApJ}
\shorttitle{On the Evolution of and Emission from GPS Sources}
\shortauthors{Stawarz et al.}

\begin{document}

\title{On the Evolution of and High-Energy Emission from GHz-Peaked-Spectrum Sources}

\author{\L . Stawarz\altaffilmark{1, 2}, L. Ostorero\altaffilmark{3, 4}, M.C. Begelman\altaffilmark{5}, R. Moderski\altaffilmark{6}, J. Kataoka\altaffilmark{7}, S. Wagner\altaffilmark{8}}

\email{stawarz@slac.stanford.edu}

\altaffiltext{1}{Kavli Institute for Particle Astrophysics and Cosmology, Stanford University, Stanford CA 94305} 
\altaffiltext{2}{Astronomical Observatory, Jagiellonian University, ul. Orla 171, 30-244 Krak\'ow, Poland}
\altaffiltext{3}{Dipartimento di Fisica Generale `Amedeo Avogadro', Universit\`a degli Studi di Torino, Via P. Giuria 1, 10125, Torino, Italy}
\altaffiltext{4}{Istituto Nazionale di Fisica Nucleare (INFN), Sezione di Torino, Via P. Giuria 1, 10125 Torino, Italy}
\altaffiltext{5}{Joint Institute for Laboratory Astrophysics, University of Colorado, Boulder, CO 80309-0440, USA}
\altaffiltext{6}{Nicolaus Copernicus Astronomical Center, Bartycka 18, 00-716 Warsaw, Poland}
\altaffiltext{7}{Department of Physics, Tokyo Institute of Technology, 2-12-1, Ohokayama, Meguro, Tokyo 152-8551, Japan}
\altaffiltext{8}{Landessternwarte Heidelberg, K\"onigstuhl, and Max-Planck-Institut f\"ur Kernphysik, Saupfercheckweg 1, Heidelberg 69117, Germany}

\begin{abstract}
Here we discuss evolution and broad-band emission of compact ($<$ kpc) lobes in young radio sources. We propose a simple dynamical description for these objects, consisting of a relativistic jet propagating into a uniform gaseous medium in the central parts of an elliptical host. In the framework of the proposed model, we follow the evolution of ultrarelativistic electrons injected from a terminal hotspot of a jet to expanding lobes, taking into account their adiabatic energy losses as well as radiative cooling. This allows us to discuss the broad-band lobe emission of young radio sources. In particular, we argue that the observed spectral turnover in the radio synchrotron spectra of these objects cannot originate from the synchrotron self-absorption process but is most likely due to free-free absorption effects connected with neutral clouds of interstellar medium engulfed by the expanding lobes and photoionized by active centers. We also find a relatively strong and complex high-energy emission component produced by inverse-Compton up-scattering of various surrounding photon fields by the lobes' electrons. We argue that such high energy radiation is strong enough to account for several observed properties of GHz-peaked-spectrum (GPS) radio galaxies at UV and X-ray frequencies. In addition, this emission is expected to extend up to GeV (or possibly even TeV) photon energies and can thus be probed by several modern $\gamma$-ray instruments. In particular, we suggest that GPS radio galaxies should constitute a relatively numerous class of extragalactic sources detected by {\it GLAST}.
\end{abstract}


\section{Introduction}

`GHz-peaked-spectrum' (GPS) objects are powerful radio sources whose spectra are inverted ($L_{\nu} \propto \nu^{-\alpha}$ with $\alpha < 0$) below peak (or turnover) frequencies $\nu_{\rm p} \sim 0.5-10$\,GHz and whose linear sizes are $LS \lesssim 1$\,kpc. `Compact steep spectrum' (CSS) objects are similarly powerful inverted-spectrum radio sources but with peak frequencies in a lower frequency range when compared to the GPS population, $\nu_{\rm p} \lesssim 0.5$\,GHz, and with larger linear sizes, $LS \sim 1-10$\,kpc. Nuclei of GPS and CSS objects can be classified as either radio galaxies, quasars, or Seyfert galaxies of type 1 or 2. Morphologically, GPS/CSS sources may be reminiscent of a smaller version of classical doubles (FR\,II radio galaxies), with pairs of symmetric lobes present at opposite sides of weak nuclei. In such cases, they are called `compact symmetric objects' (CSOs), if $LS \lesssim 1$\,kpc, or `medium symmetric objects' (MSOs), if $LS \sim 1-10$\,kpc. Quite often, however, GPS/CSS sources are characterized rather by a `core-jet' morphology with asymmetric lobes (if present at all). In such cases, it is not clear whether they should be classified as `true' GPS/CSS-es or rather as `regular' (i.e. extended) radio-loud active galactic nuclei (AGNs) viewed in projection. About $10\%$ of radio sources found in high-frequency radio surveys belong to the GPS class whereas $30 \%$ are classified as CSS objects. An extensive review on this issue was presented by \citet{odea98}.

As shown by \citet{odea97}, a relatively tight correlation between the turnover frequency and the source's linear size, namely $\log (\nu_{\rm p}/{\rm GHz}) = -0.21(\pm 0.05) - 0.65(\pm 0.05) \times \log (LS/{\rm kpc})$, holds for the investigated parameter range $\nu_{\rm p} = 0.05-20$\,GHz and $LS = 0.01-20$\,kpc. This unifies the GPS and CSS populations and suggests that they are both manifestations of the same physical phenomenon. Moreover, such a continuous distribution up to the observationally limited peak frequencies $\nu_{\rm p} \sim 10$\,GHz suggests that there may be an unnoticed population of sources with $\nu_{\rm p} > 10$\,GHz \citep{odea97,tor00}. These sources are called `high frequency peakers' (HFP), and several possible candidates were already selected \citep{dal00}. However, the majority of the candidates (especially those that are quasar-hosted) possess clear core-jet morphology and thus may be not related to the GPS/CSS phenomenon \citep{ori06}. In this context, monitoring studies may help in performing the proper classification since little variability ($< 10\%$) is expected for the discussed class of objects \citep{tor01,tin05,tor05,ori07}.

The `true' GPS/CSS sources have to be \emph{intrinsically} very powerful in radio because Doppler and projection effects seem to be, in those cases, rather marginal \citep{fan90,wil94,sai95}, with the possible exception of GPS quasars \citep{stan01,stan03}. In particular, radio powers of the considered sources at $5$\,GHz always exceed the FR\,I/FR\,II division, $L_{\rm 1.4\,GHz} \gtrsim 10^{25}$\,W\,Hz$^{-1}$, and reach $L_{\rm 1.4\,GHz} \gtrsim 10^{29}$\,W\,Hz$^{-1}$ in some cases. As shown by \citet{stan98}, $10\%-20\%$ of GPS objects possess, in addition, faint extended radio emission, with the famous 0108+388 being the most obvious example \citep{bau90,stan90}. Such extended radio halos may reach even Mpc scales \citep{sch99,mar03} and are believed to represent fossil structures formed in previous epochs of the jet activity. This idea is supported by the presence of a GPS-like radio core in the source J1247+6723, which is characterized by a classical `double-double' (i.e., restarting) large-scale radio morphology \citep{sai07}. On the other hand, as argued by \citet{stan05}, the extended emission is most often seen in GPS quasars, which are more likely core-jet like than truly compact structures and therefore not necessarily represent fossil lobes in all cases. 

Although there is an emerging agreement that GPS/CSS sources --- at least those truly compact and not simply shortened by the projection effects --- are young versions of extended radio galaxies and quasars (see \S\,2.1 below), several key questions regarding these objects remain open. They concern, for example, (i) the nature of the absorption mechanism responsible for the observed inverted radio spectra at low frequencies, (ii) details of the dynamical evolution and interaction with the ambient (galactic) medium, and also (iii) the parameters of the central engine like the accretion rate, the nuclear obscuration, etc. Clearly, a detailed analysis of the \emph{broad-band} emission from GPS sources, including recent observations in the X-ray photon energy range, may help to answer some of these questions. Here we explore the possibility that young radio galaxies may be, in addition, sources of relatively intense $\gamma$-ray emission and that detection of such radiation (or even the establishment of upper limits to it) by instruments like {\it GLAST}, {\it AGILE}, H.E.S.S., MAGIC, or VERITAS in the GeV-TeV photon energy range can help to constrain the physics of this class of objects. In particular, in \S\,2 we propose a simple and updated dynamical description of the evolution of GPS sources, which allows us to discuss in \S\,3 the expected broad-band emission of their lobes, including the GeV photon energy range. Final conclusions are given in \S\,4 of the paper. In a subsequent paper, we carefully select from the literature several GPS radio galaxies (where the lobe emission is expected to dominate the total radiative outputs, since the underlying relativistic jet and accretion disk emissions are likely to be Doppler-hidden and/or obscured) that are detected at X-ray energies and analyze their multiwavelength radiation in the framework of the presented model.

\section{Evolution of GPS Sources}

\subsection{Present Understanding}

Since GPS/CSS objects are as powerful as classical doubles but much smaller, they can be either young versions of the extended radio sources \citep{phi82} or examples of radio-loud AGNs `frustrated' by the ambient medium \citep{van84}. Efficient confinement of an expanding radio structure by a dense galactic environment was proposed to be associated in a natural way with the narrow-line region (NLR). We note that the typical parameters --- temperature, average number density, and filling factor --- of the NLR clouds, as observed in many powerful radio galaxies, are $T \sim 10^4$\,K, $n_{\rm NLR} \sim 10^3-10^4$\,cm$^{-3}$, and $\phi \sim 10^{-4}$, respectively. These NLR clouds, distributed around galactic nuclei on kpc scales with total masses up to $M_{\rm NLR} \sim 10^7 \, M_{\odot}$, are embedded within hot, X-ray emitting gaseous halos, whose typical densities (inferred from X-ray observations of giant ellipticals) are $n_{\rm ISM}(\leq 1\, {\rm kpc}) \sim 0.1$\,cm$^{-3}$. However, the frustration scenario requires total masses of cold ambient gas in a range $10^{10}-10^{11} \, M_{\odot}$ within the host galaxies \citep{dey93,car94,car98}. Such significantly denser environments of GPS/CSS sources were indeed claimed previously \citep{gop91} but are not supported by the most recent multiwavelength studies \citep[although see recently][]{gar07}.

In the framework of the youth scenario \citep{phi82}, the evolution of GPS/CSS sources toward extended FR\,IIs was followed by \citet{car85} and then by a number of authors. \citet{fan95} found that, in order to explain the size distribution of radio galaxies in this approach, one has to invoke a decrease of the radio power with increasing linear size for young, compact objects. Strong negative luminosity evolution of GPS/CSS sources \citep[advocated also by][]{rea96} would imply that most GPS/CSS objects cannot be precursors of the most luminous FR\,IIs but only of the low-power radio galaxies, located close to the FR\,II/FR\,I division. In this respect, \citet{beg96} presented a simple evolutionary model that successfully accounted for many observed features of the GPS/CSS class. In particular, assuming (i) a power-law density profile of the ambient medium $\rho_{\rm ISM}(r) \propto r^{-\beta}$ with $\beta \sim 1.5 - 2$ (where $r$ is the distance from the galactic nucleus), (ii) constant jet power during the source's lifetime, and (iii) self-similar expansion of the overpressured lobe (close to energy equipartition), \citet{beg96} obtained an almost constant advance velocity of the hotspots, a decrease of the radio power with size $L_{\rm R} \propto r^{-0.5}$, and a size distribution $dN/d \log LS \propto r^{-m+1}$ with $m \sim 0.6$ \citep[as implied by observations for $LS$ up to hundreds of kpc; see][and references therein]{fan01}. 

\citet{odea97} noted, however, a more complex size distribution of radio sources, consisting of a plateau $dN/d \log LS \propto const$ for $0.3$\,kpc $< LS <$ $10$\,kpc and a power-law tail $\propto LS^{0.4}$ for $LS > 10$\,kpc. This could imply an overabundance of compact radio sources when compared to the number of extended ones. As shown by \citet{rey97}, such an overabundance can be incorporated into the simple self-similar evolutionary model only if jet intermittency is introduced ($10^4$\,yr-long burst of jet activity recurring every $10^5$\,yrs). Alternatively, as discussed by \citet{ale00} and \citet{sne00}, an additional population of short-lived radio sources that die before reaching $>10$\,kpc scales may explain the observational results of \citet{odea97}, which, however, still needs to be confirmed within a sample of sources not affected by the projection effects \citep[but see][]{tin06}. The issue of self-similarity in the evolution of radio galaxies is still in general debated, and non-self-similar evolutionary models for compact sources, enriched by some additional effects like energy/momentum losses of the jets due to interactions with the surrounding medium \citep{dey97,per02,kaw06}, were discussed. Self-similar scenarios were also explored in more detail, enriched by (more consistent with observations) King-type ambient medium density profiles $\rho_{\rm ISM}(r) \propto (1 + (r/r_{\rm c})^2)^{-\beta/2}$ instead of a single power-law considered earlier. The presence of a plateau in the ambient gas density within the core radius $r_{\rm c} \sim 1$\,kpc implies that, in the initial state, the radio luminosity of a single source may even increase with increasing size for $LS < r_{\rm c}$ and then decrease for $LS > r_{\rm c}$ \citep{ale00,sne00}.

\subsection{A Simple Dynamical Model}

All the models describing the evolution of GPS/CSS sources start from the set of equations discussed by \citet{beg89} in the context of classical doubles expanding in an ambient medium with density profile $\rho = \rho(r)$. These equations can be derived by (i) balancing the momentum flux of a relativistic jet by the ram-pressure of the ambient medium spread over some area $A_{\rm h}$, possibly larger than the jet cross-section $L_{\rm j} / c = \rho \, v_{\rm h}^2 \, A_{\rm h}$, where $v_{\rm h}$ is the advance velocity of the jet head and $L_{\rm j}$ is the jet kinetic power; (ii) setting the lobe's sideways expansion velocity equal to the speed of the shock driven by the overpressured cocoon with internal pressure $p$ in the surrounding medium, $v_{\rm c} = (p / \rho)^{1/2}$; and (iii) assuming that all the energy transported by a pair of jets during the source's lifetime $t$ is transformed at the jet head (terminal shock) into the cocoon's internal pressure, $p \, V = 2 \, (\hat{\gamma} - 1) \, L_{\rm j} \, t$, where $V$ is the volume of the cocoon and $\hat{\gamma} = 4/3$ is the adiabatic index of the ultrarelativistic cocoon's fluid. Introducing the source linear size $LS$ and its transverse size $l_{\rm c}$, one can therefore write
\begin{eqnarray}  
L_{\rm j} = c \, \rho(LS) \, v_{\rm h}^2 \, A_{\rm h} \quad , \quad p = \rho(l_{\rm c}) \, v_{\rm c}^2 \quad , \quad 3 \, p \, V = 2 \, L_{\rm j} \, t \quad , \nonumber \\
v_{\rm h} = {d \, LS \over dt} \quad , \quad v_{\rm c} = {d \, l_{\rm c} \over dt} \quad , \quad {d \, V \over dt} = 2 \pi \, l_{\rm c}^2 \, v_{\rm h} \quad .
\end{eqnarray}
\noindent
For a given jet power $L_{\rm j}$, source linear size $LS$, and ambient medium density profile $\rho(r)$ hereafter assumed to possess King-type form, one also has to introduce some additional scaling between the model parameters. Here we follow \citet{kaw06} with $l_{\rm c}^2 \propto t^{\delta}$ and fix $\delta=1$ in order to reproduce the initial (ballistic, or `1D') phase of the jet propagation into a uniform ambient medium as found in the numerical simulations of \citet{sch02}. We also restrict our analysis to young GPS sources, which evolve in the central plateau of the galactic gaseous halo and thus have $LS < r_{\rm c} \sim 1$\,kpc. Therefore, we set the ambient density profile as $\rho = m_{\rm p} \, n_0$ with $n_0 \approx 0.1$\,cm$^{-3}$ \citep[see][]{mat03}. Such a choice gives $v_{\rm h} \propto LS^0$, $v_{\rm c} \propto LS^{-1/2}$, $p \propto LS^{-1}$, $l_{\rm c} \propto LS^{1/2}$, $V \propto LS^{2}$, $t \propto LS$, and $A_{\rm h} \propto LS^0$. Note, that the source linear size $LS$ is defined here as the jet length, i.e., as the separation of the terminal hotspot from the core.

In order to fix some other model parameters, we recall here several observational findings. \citet{kat97} analyzed the radio emission of the two CSS sources 3C\,67 and 3C\,190 and found that the spectra of their lobes are consistent with relatively young source ages of $t \sim 10^4-10^5$\,yr for the equipartition magnetic fields of $B \sim 1$\,mG. This implies that the hotspots' advance velocities, $v_{\rm h} \sim 0.3c$, are significantly (an order of magnitude) higher than the analogous values found for classical doubles. \citet{mur99} performed spectral ageing studies for a number of other CSS sources and found that, assuming again energy equipartition, the resulting source ages are indeed $< 10^5$\,yr and imply the average advance velocities $v_{\rm h} \sim 0.3 c$. Such high velocities were in fact detected directly, first in GPS sources 0710+439 \citep{ows98a} and 0108+388 \citep{ows98b}. Since then, several radio observations of hotspots in many GPS/CSO objects have confirmed repeatedly that $v_{\rm h} \sim 0.2 \, h^{-1} c$ \citep{tay00,tsc00,gir03,pol03,gug05,nag06,gug07,luo07}. This value should, in fact, be divided by a factor of 2, because typically in the literature the relative hot-spot separation is discussed, i.e., the separation of one hot-spot relative to the other, and not the separation of the hotspot relative to the very faint (and therefore often undetected) core. In general, the established agreement between kinematic and spectral ages for the GPS and CSS populations ($< 10^4$\,yr for GPS and $\sim 10^4-10^5$\,yr for CSS objects) supports approximately the fulfillment of energy equipartition within their lobes, at least in the majority of studied sources (magnetic fields $\sim 10$\,mG and $\sim 1$\,mG for GPS and CSS classes, respectively). In addition, it implies --- by means of ram-pressure arguments --- an ambient gaseous/interstellar medium (ISM) density $n_{\rm ISM} \lesssim 1$\,cm$^{-3}$ on scales between a few and a few hundred pc (in agreement with the value anticipated here).

Constant advance velocity of GPS/CSS objects in the simple model presented here, $v_{\rm h} \propto LS^0$, is in good agreement with observations, and thus, we fix hereafter $v_{\rm h} \approx 0.1 c$. Another constraint is provided by the established (approximate) minimum power condition. In general, one can parameterize the magnetic field energy density in the expanding lobes as $U_{\rm B} = \eta_{\rm B} \, p$, with $\eta_{\rm B} < 3$. Hence, with the model pressure
\begin{equation}
p = \left({L_{\rm j} \, m_{\rm p} n_0 \, v_{\rm h} \over 6 \pi}\right)^{1/2}\!\!LS^{-1} \approx 5.3 \times 10^{-7} \, L_{\rm j, \, 45}^{1/2} \, LS_{100}^{-1} \, {\rm erg \, cm^{-3}} ,
\end{equation}
\noindent
where $LS_{100} \equiv LS/100$\,pc and $L_{\rm j, \, 45} \equiv L_{\rm j} / 10^{45}$\,erg\,s$^{-1}$, the magnetic field intensity within the lobes is expected to scale like
\begin{equation}
B = \left( 8 \pi \, \eta_{\rm B} \, p\right)^{1/2} \approx 3.6 \, \eta_{\rm B}^{1/2} \, L_{\rm j, \, 45}^{1/4} \, LS_{100}^{-1/2} \, {\rm mG} \, .
\end{equation}
\noindent
This is consistent with the equipartition values $B \sim 1-10$\,mG typically obtained for the CSS ($LS \sim 1-10$\,kpc) and GPS ($LS \sim 0.01-1$\,kpc) objects, if $\eta_{\rm B}^{1/2} L_{\rm j, \, 45}^{1/4} \gtrsim 1$. Note the comfortably weak dependence of the model magnetic field $B$ on the jet kinetic power $L_{\rm j}$. Note also that the agreement between the ages of GPS sources derived by means of the spectral ageing analysis and the dynamical one is in fact expected in the presented model since, in its framework, the age of the source during the GPS phase of the evolution is simply
\begin{equation}
t = v_{\rm h}^{-1} \, LS \approx 3.3 \times 10^3 \, LS_{100} \, {\rm yrs} \, .
\end{equation}
\noindent
The other model parameters can be evaluated as
\begin{eqnarray}  
A_{\rm h} = \left({L_{\rm j} \over c \, m_{\rm p} n_0 \, v_{\rm h}^2}\right) \approx 2.2 \times 10^{40} \, L_{\rm j, \, 45} \, {\rm cm^2} \, ,\\ 
l_{\rm c} = \left({8 \, L_{\rm j} \over 3 \pi \, m_{\rm p} n_0 \, v_{\rm h}^3}\right)^{1/4}\!\!LS^{1/2} \approx 3.7 \times 10^{20} \, L_{\rm j, \, 45}^{1/4} \, LS_{100}^{1/2} \, {\rm cm} \, ,\\ 
V = \left({8 \pi \, L_{\rm j} \over 3 \, m_{\rm p} n_0 \, v_{\rm h}^3}\right)^{1/2}\!\!LS^{2} \approx 1.3 \times 10^{62} \, L_{\rm j, \, 45}^{1/2} \, LS_{100}^{2} \, {\rm cm^3} \, ,\\
v_{\rm c} = \left({L_{\rm j} \, v_{\rm h} \over 6 \pi \, m_{\rm p} n_0}\right)^{1/4}\!\!LS^{-1/2} \approx 1.8 \times 10^{9} \, L_{\rm j, \, 45}^{1/4} \, LS_{100}^{-1/2} \, {\rm cm \, s^{-1}} \, .
\end{eqnarray}
\noindent

Interestingly, the sideways expansion velocity $v_{\rm c}$, again weakly dependent on the jet power, is rather high but consistent with the outflow velocities of the line-emitting gas (believed to be pushed out and accelerated by the expanding lobes), $v_{\rm out} \gtrsim 10^8$\,cm\,s$^{-1}$, observed at optical frequencies (with luminosities of about $\sim 10^{42}-10^{43}$\,erg\,s$^{-1}$ for individual lines) in many CSS objects \citep{dev99,odea02}. In addition to this, gaseous outflows in host galaxies of GPS/CSS sources may be manifested as blue-shifted absorption-line systems at UV frequencies. This, in fact, was observed in the CSS quasar 3C\,48 \citep{gup05}, indicating outflow velocities of, again, $v_{\rm out} \sim 10^8$\,cm\,s$^{-1}$, driven by interaction of the expanding jets/lobes with the kpc-scale gaseous environment. We note that, since the $v_{\rm c}$ evaluated here may be higher than the expected sound speed in the external medium, $c_s = (5 k T / 3 m_{\rm p})^{1/2} \approx 3.7 \times 10^7$\,cm\,s$^{-1}$ (for the anticipated temperature $T \approx 10^7$\,K characterizing the hot phase of the gaseous environment), a bow shock may be expected to form around radio lobes of GPS sources, with a possibly high Mach number $\mathcal{M}_{\rm sh} = (3 p / 5 n_0 k T)^{1/2} \approx 48 \, L_{\rm j, \, 45}^{1/4} \, LS_{100}^{-1/2}$.

\subsection{Synchrotron Emission}

\subsubsection{Synchrotron Luminosity Evolution}

With the electron energy distribution injected from the terminal jet shock to the expanding lobe and modifid thereby by the adiabatic and radiative cooling effects, $N_{\rm e}(\gamma)$, one can express the lobes' synchrotron luminosity as $L_{\rm syn} = \left(4 \, c \, \sigma_{\rm T} / 3 \, m_{\rm e} c^2 \right) \, f\!(\gamma) \, U_{\rm B} \, U_{\rm e} \, V$, where $f\!(\gamma) \equiv \langle \gamma^2 \rangle / \langle \gamma \rangle = \int \gamma^2 \, N_{\rm e}(\gamma) \, d\gamma / \int \gamma \, N_{\rm e}(\gamma) \, d\gamma$ and the electron energy density is simply $U_{\rm e} = m_{\rm e} c^2 \, \int \gamma \, N_{\rm e}(\gamma) \, d\gamma$. Assuming further that $U_{\rm e} = \eta_{\rm e} \, p$ with $\eta_{\rm e} \lesssim 3$ (i.e., that the jet electrons shocked at the terminal hotspot, possibly in rough equipartition with the magnetic field and relativistic protons, provide the bulk of the lobes' pressure) and that the lobes' electron population does not change significantly its spectral shape during the GPS phase of the expansion (see \S\,3.1), the synchrotron luminosity turns out to be constant with time and independent of the source linear size, $L_{\rm syn} \propto U_{\rm B} \, U_{\rm e} \, V \propto LS^0$, in particular
\begin{eqnarray}  
L_{\rm syn} & = & {4 \sigma_{\rm T} \over 9 m_{\rm e} c} \, \left({2 m_{\rm p} n_0 \over 3 \pi \, v_{\rm h}}\right)^{1/2}\!\!\eta_{\rm B} \, \eta_{\rm e} \, f\!(\gamma) \, L_{\rm j}^{3/2} \approx \nonumber \\
& \approx & 1.2 \times 10^{42} \, \eta_{\rm B} \, \eta_{\rm e} \, f\!(\gamma) \, L_{\rm j, \, 45}^{3/2} \, {\rm erg \, s^{-1}} \, .
\end{eqnarray}
\noindent
Such an evolution is expected to hold only if the source is not older than $10^5$\,yr (i.e., the jet is in the initial, ballistic evolution phase) and if the ambient medium density profile can be approximated as being constant (i.e., if $LS \lesssim 1$\,kpc, the typical core radius of the gaseous medium in giant ellipticals). However, for $t > 10^5$\,yr, the above scaling breaks down, and the synchrotron luminosity is expected to decrease with increasing linear size of the source \citep[see the discussion in][]{beg96,kaw06}. Note that equation 9 implies the interesting constraint $1 \leq \eta_{\rm B} \, \eta_{\rm e} \, L_{\rm j, \, 45}^{3/2} \, f\!(\gamma) \leq 10^4$ on the presented model because the observed radio luminosities of GPS sources (generally selected so far at intermediate/high flux density) are in the range $L_{\rm syn} \sim 10^{42} - 10^{46}$\,erg\,s$^{-1}$.

The scaling of $L_{\rm syn}$ given above corresponds to a constant electron injection provided by the terminal hotspot, no absorption of (radio) synchrotron photons within the lobes, and to the fixed spectral function $f(\gamma)$ during the source's evolution (see in this context \S\,3.1). Note also that the monochromatic synchrotron power produced by electrons with a given Lorentz factor $\gamma^{\star}$, for which absorption effects can be again neglected, goes like $[\nu L_{\nu}]_{\nu \propto {\gamma^{\star}}^2} \propto U_{\rm B} \, U_{\rm e} \, V \propto LS^0$. On the contrary, the monochromatic synchrotron power measured at a fixed observed frequency $\nu^{\star}$ (and thus produced by the electrons with different energies at different evolutionary stages, due to a change in the lobes' magnetic field intensity) scales as $[\nu^{\star} L_{\nu^{\star}}] \propto B^{(s-3)/2} \, U_{\rm B} \, U_{\rm e} \, V \propto LS^{(3-s)/4}$ for a power-law electron energy $N_{\rm e}(\gamma) \propto \gamma^{-s}$ since the monochromatic luminosity can be written as
\begin{eqnarray}  
[\nu L_{\nu}]_{\rm syn} & = & {2 \over 3} c \, \sigma_T \, V \, U_{\rm B} \, \left[ \gamma^3 \, N_{\rm e}(\gamma) 
\right]_{\gamma = \sqrt{4 \, \pi \, m_{\rm e} c \, \nu / 3 \, e \, B}} \nonumber \\
& \rightarrow & {2 \, c \, \sigma_T \over 3\, m_{\rm e} c^2} \, {V \, U_{\rm B} \, U_{\rm e} \over \int \gamma^{1-s} \, d\gamma } \, \left({4 \, \pi \, m_{\rm e} c \, \nu \over 3 \, e \, B}\right)^{(3-s)/2} \quad {\rm for} \quad N_{\rm e}(\gamma) \propto \gamma^{-s} \, .
\end{eqnarray}
\noindent

\subsubsection{Absorption Effects}

The observed turnover in the radio spectra of GPS/CSS sources is their main characteristic. It was proposed that it is due to either synchrotron self-absorption (SSA) or free-free absorption (FFA) by an inhomogeneous screen of dense ambient matter. The observed spectral indices below the peak frequency are usually $\alpha_{\rm low} \geq -2$; in some cases, they are close to the standard value $-5/2$ predicted by the homogeneous SSA model while in other cases, they are even consistent with the exponential cutoff predicted by the simplest version of the FFA model. The variety of the low-frequency spectral indices thus indicates inhomogeneity of the absorbing medium and/or superposition of several emission components with different physical parameters. \citet{dev97} found that the average spectral indices for the analyzed sample of GPS/CSS sources are $\alpha_{\rm low} = -0.51 (\pm 0.03)$ and $\alpha_{\rm high} = +0.73 (\pm0.06)$ below and above the peak frequency, respectively and that, in addition, the values of $\alpha_{\rm high}$ are characterized by a very broad distribution between $+0.5$ and $+1.2$. They also claimed a flat spectral plateau between $\nu_{\rm p}$ and $2 \times \nu_{\rm p}$ in the template GPS/CSS spectrum, with average power-law slope $\alpha_{\rm peak} = +0.36 (\pm 0.05)$, ascribed to the broadening of the spectral peaks by the sources' substructure.

An implication of the evolutionary models for GPS/CSS sources with strong negative luminosity evolution \citep[i.e., the models assuming steep ambient medium density profile $\rho_{\rm ISM}(r) \propto r^{-\beta}$ with $\beta >1$, like that of][]{beg96} is that the young sources evolve on the $\nu_{\rm p}-LS$ plane as $\nu_{\rm p} \propto LS^{-x}$ with $x>1$ if the spectral turnover is due to the SSA process. Thus, the GPS/CSS sources do not evolve along the observed $\nu_{\rm p} \propto LS^{-0.65}$ line but `leave' the $\nu_{\rm p}-LS$ plane when their radio powers decrease enough with increasing $LS$ \citep{odea97}. This could possibly explain the observed scatter in the $\nu_{\rm p}-LS$ correlation. On the other hand, \citet{bic97} successfully reproduced the observed $\nu_{\rm p} \propto LS^{-0.65}$ dependence in the framework of a model in which the spectral turnover is due to FFA by a clumpy/filamentary multi-phase ISM, modified (ionized) by the passage of a bow shock due to expanding radio lobes \citep[see also in this context][]{kun98}. However, this model requires very special parameters for the ambient medium (its high density, in particular) not consistent with the most recent observations for the majority of sources \citep[see the discussion in][]{beg99}. The other promising possibility left is therefore the `engulfed cloud' scenario proposed by \citet{beg99}, in which the neutral clouds of ISM penetrating the expanding radio lobe and photoionized by the nuclear radiation are responsible for the spectral turnover of GPS sources due to free-free absorption of the radio photons. 

Can the SSA effects be responsible for the observed spectral turnover of young radio sources in the framework of the dynamical model proposed here? To investigate this, we note that the characteristic SSA frequency can be found from the equation $\tau_{\nu}^{\rm ssa} =1$, where $\tau_{\nu}^{\rm ssa} = \kappa_{\nu}^{\rm ssa} \, LS$ is the optical depth for the synchrotron self-absorption process described by the absorption coefficient $\kappa_{\nu}^{\rm ssa}$ within the uniform medium of the spatial scale $LS$. Since, in the case of a power-law electron energy distribution, this coefficient is $\kappa_{\nu}^{\rm ssa} \propto N_0 \, B^{(s+2)/2} \, \nu^{-(s+4)/2}$, the characteristic (peak) SSA frequency is $\nu_{\rm ssa} \propto LS^{-x}$, with $x = (s+2)/(2s +8) = 0.3 - 0.36$ for $s = 1-3$. This is flatter than the observed distribution of the turnover frequency in GPS/CSS sources, $\nu_{\rm p} \propto LS^{-0.65}$. On the other hand, as noted before, the observed $\nu_{\rm p}-LS$ distribution may be shaped by some additional factors related to the sources' luminosity evolution and to the incompleteness of the samples considered and thus may not reflect directly the evolution of the peak frequency in a single source. Note, in this context, that in the framework of the discussed model (see equation 10) the synchrotron power at such an SSA frequency scales as $[\nu_{\rm ssa} L_{\nu_{\rm ssa}}] \propto B^{(s-3)/2} \, \nu_{\rm ssa}^{(3-s)/2} \, U_{\rm B} \, U_{\rm e} \, V \propto LS^{y}$, with $y = 0.2 - 0$ for $s = 1-3$; i.e., it is expected either to increase slightly or even to remain constant with increasing linear size of the source. Thus, it seems unlikely that the evolving young sources `leave' quickly the flux-limited $\nu_{\rm p}-LS$ plane due to a decrease in their peak luminosity \citep[cf.][]{odea97}.

In addition, SSA effects are expected to manifest themselves at relatively low frequencies. To illustrate this, let us assume that the initial electron energy spectrum injected from the hotspots to the expanding lobes is, on average, of a power-law form with the `standard' spectral index $s = 2$ (although the obtained results hold approximately for a broader range of $1 < s < 3$). In such a case, one gets $U_{\rm e} = m_{\rm e} c^2 \, N_0 \, \ln[\gamma_{\rm max}/\gamma_{\rm min}] \approx 10 \, m_{\rm e} c^2 \, N_0$, and
\begin{equation}
\kappa_{\nu}^{\rm ssa} = 0.148 \, {3 e^4 \over 4 \pi \, m_{\rm e}^3 c^3} \, N_0 \, B^2 \, \nu^{-3} \, .
\end{equation}
\noindent
Hence, the condition $\tau_{\nu}^{ssa} = 1$ gives the critical SSA frequency
\begin{eqnarray}  
\nu_{\rm ssa} & = & \left({0.148 \, e^4 \, m_{\rm p} \, n_0 \, v_{\rm h} \, \eta_{\rm e} \, \eta_{\rm B} \, L_{\rm j} \over 10 \, \pi \, m_{\rm e}^4 c^5 \, LS }\right)^{1/3} \approx \nonumber \\
& \approx & 0.3 \, \eta_{\rm e}^{1/3} \, \eta_{\rm B}^{1/3} \, L_{\rm j, \, 45}^{1/3} \, LS_{100}^{-1/3} \quad {\rm GHz} \, .
\end{eqnarray}
\noindent
Because $\eta_{\rm B}, \, \eta_{\rm e} \lesssim 3$ and the expected jet kinetic power is $L_{\rm j} < 10^{47}$\,erg\,s$^{-1}$ (see below), this frequency is much lower than the turnover frequency observed in GPS/CSS sources, allowing us to neglect SSA effects in the following discussion.

The above conclusion is in agreement with several observational supports for the free-free absorption process shaping the inverted spectra of GPS/CSS objects. For example, \citet{pec99}, \citet{kam00}, and \citet{mar01} argued for FFA in GPS sources 1946+708, OQ\,208 and 0108+388, respectively, based on investigations of spectral index maps and hence on differences between low-frequency spectra in different parts of the sources. Their modeling implies a non-uniform absorbing gaseous medium with high average number density ($\sim 10^3$\,cm$^{-3}$), ascribed by the authors to, e.g., a clumpy torus-like structure of the obscuring material with extension $\lesssim 100$\,pc. Supporting this interpretation, \citet{kam03} showed that, on average, type 1 (Seyfert 1, quasars) and type 2 (Seyfert 2, radio galaxies) GPS sources have different characteristics of the absorption features (asymmetric and symmetric distributions with respect to their nuclei, respectively), consistent with the FFA process associated with anisotropically distributed obscuring material. Finally, \citet{mut02} noted that the SSA model implies a change of the polarization angle by $90^{\circ}$ across the spectral peak, i.e., between the optically-thick and optically-thin parts of the continuum. Since no such changes were observed in the sample considered, \citet{mut02} argued for FFA effects playing the major role. One should be aware, however, that the GPS quasars may not be in reality compact/young sources, as emphasized in \S~1 of this paper, and therefore that they may be intrinsically different from the truly young GPS radio galaxies.

Let us therefore discuss in more detail the absorption model proposed by \citet{beg99}. In this model, dense hydrogen clouds of ISM present at pc$-$kpc distances from the centers of young radio sources and engulfed by their expanding lobes are photoionized by the active nuclei, causing free-free absorption of the lobe radio emission. Such clouds may be naturally identified with the ones producing narrow-line emission and also H{\sc i} absorption lines. Very broad absorption lines are indeed often detected in GPS/CSS objects (at much higher rates than in extended radio galaxies) with neutral hydrogen column densities from $N_{{\rm H{\sc I}}} \lesssim 10^{22}$\,cm$^{-2}$ down to $N_{{\rm H{\sc I}} } \gtrsim 10^{19}$\,cm$^{-2}$ \citep{ver03,pih03,gup06}. The evaluated column densities were claimed to anticorrelate with the source sizes, $N_{{\rm H{\sc I}}} \propto LS^{-0.45}$ \citep{pih03,gup06}. Previously, it was speculated that the appropriate absorbing medium possesses a torus-like distribution. However, detailed studies of a few objects performed to date \citep{lab06,ver06} indicate that the H{\sc i} absorption lines are associated with optical emission lines and thus arise most likely in the atomic cores of NLR clouds interacting with the expanding radio source. Interestingly, the intensity ratios of lines produced by spatially resolved NLR clouds in nearby Seyfert galaxies imply a density decrease with distance from the ionizing source (galactic center) $\propto r^{-n}$, with $1 < n < 2$ \citep[see][]{kai00,nel00,kra00a,kra00b,mun03,bra04}. This would then be consistent with the noted anticorrelation of $N_{{\rm H{\sc I}}}$ with $LS$ if H{\sc i} absorption is indeed due to NLR clouds and not to the hot phase of the ISM gas \citep[which then may obey a King-type distribution with central plateau $n_{\rm ISM} \propto const$, as assumed in this paper; cf.][]{pih03}. 

It is also interesting to note that GPS sources exhibit in general (very) low polarization of their radio fluxes \citep[on the level of a few percent, if any;][]{odea91}, while radio continua of CSS objects are polarized a bit more strongly \citep{stan01}. Such low polarization is most probably due to Faraday effects. The rotation measure observed has a very broad scatter in the GPS/CSS sample, from very large, $RM \gtrsim 10^4$\,rad\,m$^{-2}$ \citep{nan00}, to very small, $RM \lesssim 10^2$\,rad\,m$^{-2}$. In some cases, $RM$ is very different for the lobe and the counterlobe in a single source, which indicates large asymmetries in the gaseous environment on kpc-scales \citep{jun99}. Faraday screens seem to be associated predominantly with the optical line-emitting clouds interacting with jets \citep{cot03,fan04,cot06}. Indeed, the jets in CSS quasars exhibit more complex and distorted radio morphologies than the jets in flat-spectrum radio quasars \citep{man98}, suggesting significant jet-ambient medium interactions taking place somewhere within the NLR. Also, some other morphological and polarization properties often suggest an asymmetric distribution of the ambient medium surrounding CSS objects, possibly resulting in an intrinsically asymmetric structure of their lobes and jets \citep{sai95,sai01,sai03}.

In a scenario where the engulfed photoionized NLR clouds are responsible for the H{\sc i} absorption lines and the distortion of the radio structures but are unable to confine or frustrate the jets \citep{beg99}, the dominance of FFA effects by those clouds in generating the spectral turnover in the radio spectra of GPS and CSS objects would imply a phenomenological relation between the peak frequency and the source's linear size, $\nu_{\rm p} \approx 2.7 \, LS_{100}^{-0.65}$\,GHz. The synchrotron luminosity at the peak (turnover) frequency is thus expected to scale with the size of the source as $[\nu_{\rm p} L_{\nu_{\rm p}}] \propto B^{(s-3)/2} \, \nu_{\rm p}^{(3-s)/2} \, U_{\rm B} \, U_{\rm e} \, V \propto LS^{y}$, with $y = (-0.15) - 0$ for $s = 1-3$ (equation 10); i.e., it should decrease only very slightly with increasing $LS$ or even remain constant during the GPS evolution phase. Note that the low-frequency radio continua are then expected to be of the form $L_{\nu < \nu_{\rm p}} \propto \nu^{2-(s - 1)/2}$ because, in the described engulfed-cloud model, the appropriate absorption coefficient scales as $\kappa_{\nu}^{\rm ff} \propto \nu^{-2}$ \citep{beg99}. Therefore, the observed scatter in radio spectral indices below the spectral turnover may result not only from the non-uniformity of the lobes and NLR but also from the internal scatter in the low-energy electron spectral index $s$ in the lobes of GPS objects. 

\section{Broad-Band Spectra}

\subsection{Electron Energy Distribution}

It is typically assumed that the electron energy distribution formed at the jet terminal shock and injected to the lobes is of a simple power-law form $Q(\gamma) \propto \gamma^{-s}$. Such an injected spectrum undergoes further radiative and adiabatic energy losses within the expanding cocoon. For the GPS-phase of the evolution described in \S\,2.2, the lobes' expansion is given by $V/V_0 = \pi \, l_{\rm c}^2 / A_{\rm h} \propto LS$, and the cooling (magnetic and radiative fields) changes with the source size $LS$. Assuming that synchrotron emission dominates the radiative losses (as is the case for all but the most compact sources, as discussed in the next section \S\,3.2), the resulting electron energy distribution for a given source linear size $LS = v_{\rm h} \, t$ with $U_{\rm B} \propto LS^{-1}$ can be found from equation A8 of Appendix A, which reads
\begin{equation}
\mathcal{N}_{\rm e}(\gamma) = {\gamma^{-2} v_{\rm h}^{-1}\over LS^{1/3}}\!\!\int^{LS}_{LS_0}\!\!\!Q[\gamma(LS')] \, \gamma(LS')^2 {LS'}^{1/3} dLS',
\end{equation}
\noindent
where $\mathcal{N}_{\rm e}(\gamma) = N_{\rm e}(\gamma) \, V$. In the integral above,
\begin{equation}
\gamma(LS') \approx {c_2 \, \gamma_0 \, {LS'}^{-1/3} \over 1 + c_3 \, \gamma_0 \, \left(LS_0^{-1/3} - {LS'}^{-1/3}\right)}
\end{equation}
\noindent
(see equation A6), where the constants $c_2 \approx 2.5 \times 10^6 \, L_{\rm j, \, 45}^{1/6}$\,cm$^{1/3}$ and $c_3 \approx 1.34 \times 10^4 \, \eta_{\rm B} \, L_{\rm j, \, 45}^{2/3}$\,cm$^{1/3}$ follow from the equations 2, 5, and 6. For a given power-law source function $Q(\gamma) = K_{\rm e} \gamma^{-s}$ with $s>2$, one can obtain further
\begin{eqnarray}  
\mathcal{N}_{\rm e}(\gamma) & = & {3 \, v_{\rm h}^{-1} \, \gamma^{-s} \, K_{\rm e} \, LS \, \omega \over s \, (s+1) \, (s-1) \, (s+2)} \times \nonumber \\ 
& & \times \left\{ {s \, (s^2-1) \over (\omega + 1)} + {3 \, s \, (s-1) \over (\omega + 1)^2} + {6 \, (s-1) \over (\omega + 1)^3} + {6 \over (\omega + 1)^4}\right\} \, ,
\end{eqnarray}
\noindent
where, after integrating, $\gamma_0$ was replaced back with $\gamma \, LS^{1/3} / \left[c_2 + c_3 \, \gamma \, \left(1- (LS/LS_0)^{1/3}\right) \right]$, and
\begin{equation}
\omega = {c_2/c_3 \over \gamma} \equiv {\gamma_{\rm cr} \over \gamma} \approx 200 \, \eta_{\rm B}^{-1} \, L_{\rm j, \, 45}^{-1/2} \, \gamma^{-1} \, . 
\end{equation}
\noindent

Equation 15 implies that the electron energy distribution in the lobes of GPS sources, in the case of a single power-law injection $\propto \gamma^{-s}$ with $s>2$, is expected to be of a broken power-law form,
\begin{equation}
N_{\rm e}(\gamma) = {\mathcal{N}_{\rm e}(\gamma) \over V} \approx {K_{\rm e} \, LS \over v_{\rm h} \, V} \times \left\{ \begin{array}{ccc} \gamma^{- s} & {\rm for} & \gamma < \gamma_{\rm cr} \\
\gamma_{\rm cr} \, \gamma^{- s - 1} & {\rm for} & \gamma > \gamma_{\rm cr} \end{array} \right. \, .
\end{equation}
\noindent
Note that the critical break $\gamma_{\rm cr}$ does not depend on the linear size $LS$ and hence neither on the function $f\!(\gamma) = \int \gamma^2 \, N_{\rm e}(\gamma) \, d\gamma / \int \gamma \, N_{\rm e}(\gamma) \, d\gamma \approx \gamma_{\rm cr}^{3-s}/(3-s)$. In addition, since $V \propto LS^2$ (equation 7), the normalization of the electron energy distribution for a given $LS$ scales as $LS^{-1}$. Similarly, the electron energy density $U_{\rm e} = m_{\rm e} c^2 \, \int \gamma \, N_{\rm e}(\gamma) \, d\gamma \propto LS^{-1}$, ensuring that the ratio $U_{\rm e}/U_{\rm B}$ is constant during the GPS phase of the lobes' evolution. These scaling relations validate the discussion presented in \S\,2.3 and allow one to fix the normalization of the electron injection function, $K_{\rm e}$, through the assumed relation $U_{\rm e} = \eta_{\rm e} \, p$. One may finally find that for a single power-law injection with $s \sim 2 - 3$ considered in this paragraph, the total (unabsorbed) synchrotron luminosity of GPS radio galaxies is expected to be roughly $0.1 \, L_{\rm j} \lesssim L_{\rm syn} \lesssim L_{\rm j}$ (see equation 9).

In the discussion above we have assumed the injection of a single power-law electron energy distribution within the terminal hotspots of GPS sources. However, some recent studies indicate that the situation may be more complicated. For example, \citet{sta07} found that in the case of the archetype FR~II radio galaxy Cygnus A, the electron spectrum can be approximated as a broken power-law $Q(\gamma) \propto \gamma^{- s_1}$ for $\gamma < \gamma_{\rm int}$, and $Q(\gamma) \propto \gamma^{- s_2}$ for $\gamma_{\rm int} < \gamma$, with $\gamma_{\rm int} \approx m_{\rm p}/m_{\rm e}$, $s_1 \sim 1.5$ and $s_2 \gtrsim 3$. Such a form is, in fact, expected in the case of cold protons carrying the bulk of the jets' energy in powerful radio sources (at least on $>$\,pc scales) due to the nature of the particle acceleration process taking place at the mildly relativistic terminal shocks dynamically dominated by the protons \citep[see the discussion in][]{sta07}. If this is also the case for young radio sources, then one can expect the electron energy distribution within the lobes of young radio galaxies, which can be evaluated as
\begin{equation}  
N_{\rm e}(\gamma) = \gamma^{-2} \, {LS \over v_{\rm h} \, V} \, \int_{c_2^3/LS}^1\!\!Q\!\!\left[{\omega \, \gamma \over (\omega+1) \, x^{1/3} - 1}\right] \, \left({\omega \, \gamma \over (\omega +1) \, x^{1/3} - 1}\right)^2 \, x^{1/3} \, dx
\end{equation}
\noindent
(see equation 13), to be peaked on the $\gamma^2 N_{\rm e}(\gamma)-\gamma$ plane around the electron energy $\gamma_{\rm cr}$, as long as $s_1 < 2$. Hence, one gets $f\!(\gamma) \sim \gamma_{\rm cr}$, and therefore, the expected total synchrotron luminosity $L_{\rm syn} \gtrsim 0.1 \, \eta_{\rm e} \, L_{\rm j}$, similarly to the case of a single power-law injection discussed above. Thus, the proposed model with the anticipated $\eta_{\rm e} \gtrsim 1$ is in agreement with the observed values of $L_{\rm syn} \sim 10^{42}-10^{46}$\,erg\,s$^{-1}$ for different injection conditions and a jet luminosity in the range $L_{\rm j} \sim 10^{43}-10^{47}$\,erg\,s$^{-1}$. This range is indeed as expected if GPS sources are progenitors of FR~I \emph{and} FR~II radio galaxies and implies that $\gtrsim 10\%$ of the jet kinetic power is dissipated for the synchrotron emission of the GPS lobes \citep[as assumed by][]{dey93}.

The evolution of the electron energy distribution within the lobes of GPS sources, as given by equation 18, is shown in Figure~1 for different jet powers ($L_{\rm j} = 10^{44}$, $10^{45}$, $10^{46}$, and $10^{47}$\,erg s$^{-1}$), and different source linear sizes ($LS = 100$\,pc and $1$\,kpc; thick/upper and thin/lower lines, respectively). In the figure, two different source functions are considered for illustration, namely single power-law $Q(\gamma) \propto \gamma^{-2.5}$ (dotted lines), or broken power-law $Q(\gamma) \propto \gamma^{-1.5}$ for $\gamma < \gamma_{\rm int}$ and $Q(\gamma) \propto \gamma^{-3}$ for $\gamma > \gamma_{\rm int}$ (solid lines). In both cases the minimum and maximum electron Lorentz factors are $\gamma_{\rm min} = 1$ and $\gamma_{\rm max} = 10^5$, while the normalization of the injection function is evaluated through the condition $U_{\rm e} = \eta_{\rm e} \, p$ with $\eta_{\rm e} = 3$ and $\eta_{\rm B} = 0.3$. Vertical dotted and dashed lines in the figure indicate critical electron energies $\gamma_{\rm cr} = 200 \, \eta_{\rm B}^{-1} \, L_{\rm j, \, 45}^{-1/2}$ and $\gamma_{\rm int} = m_{\rm p} / m_{\rm e}$, respectively. As shown, the normalization of the electron energy distribution decreases with increasing source linear size $LS$ roughly as $LS^{-1}$ while the spectral continuum steepens at $\gamma > \gamma_{\rm cr}$ when compared to the injected one. Note also that, in the case of high jet luminosities and small source linear sizes, the peak in the electron energy distribution for a broken power-law injection may be slightly higher than $\gamma_{\rm cr}$. Thus, the total synchrotron luminosity may exceed the value $0.1 \, \eta_{\rm e} \, L_{\rm j}$ for the most powerful and compact GPS sources.

A more complex form of the electron source function $Q(\gamma)$ introduces several other interesting features in the model. In particular, with an intrinsically broken electron spectrum, further modified by the adiabatic and radiative energy losses, one should expect to observe a curved or multiply broken synchrotron continuum. That is because the synchrotron spectrum in such a case is shaped by three different critical frequencies: one related to absorption effects, $\nu_{\rm p} \approx 2.7 \, LS^{-0.65}$\,GHz, and the other two related to the critical electron energies discussed above, namely $\nu_{\rm cr} = 3 \, e B \, \gamma_{\rm cr}^2 / 4 \pi \, m_{\rm e} c \approx 0.54 \, \eta_{\rm B}^{-3/2} \, L_{\rm j, \, 45}^{-3/4} \, LS_{100}^{-1/2}$\,GHz, and $\nu_{\rm int} = 3 \, e B \, \gamma_{\rm int}^2 / 4 \pi \, m_{\rm e} c \approx 70 \, \eta_{\rm B}^{1/2} \, L_{\rm j, \, 45}^{1/4} \, LS_{100}^{-1/2}$\,GHz. All these are interestingly clustered around GHz frequencies; some spectral plateau around this range is thus to be expected. This kind of a spectral feature is indeed observed in many particular objects and is also present in a template GPS spectrum derived by \citet{dev97}.

\subsection{Photon Fields Within Lobes}

At near-UV frequencies, GPS/CSS sources, like classical doubles, exhibit complex spectra composed of nebular continuum, nuclear light (both direct and scattered), and a starburst component \citep{tad02,dev07}. An additional non-thermal contribution from jets and compact lobes was also anticipated \citep{beg99}. As suggested by \citet{lab08}, the observed \emph{extended} UV luminosities of several GPS/CSS objects, being in the range $\sim 10^{40}-10^{42}$\,erg\,s$^{-1}$, are correlated neither with the sources' radio powers nor with their linear sizes. \citet{lab08} argued that this emission, produced predominantly by relatively young stars, results from bursts of star formation that took place at the time of or before the formation of compact radio structures. Moreover, the detected (in some cases) UV component aligned with the radio axis may also indicate star formation enhanced or triggered by the expanding radio lobes. Here we concentrate on the UV photons provided directly by the active center, assuming that nuclei of GPS objects are intrinsically similar to the ones observed in quasars and Seyfert galaxies. Therefore, we assume that the bulk of the radiative output of the optically thick accretion disk is emitted at UV frequencies (thus forming the characteristic `big blue bump') very close to the central engine, with intrinsic luminosities of the order of $L_{\rm UV} \sim (10^{45}-10^{47})$\,erg\,s$^{-1}$ \citep[see, e.g.,][]{kor99} and thus dominating over the other, extended UV photon fields mentioned above. Such strong emission is in fact observed directly in many GPS quasars \citep[also in prep.]{sie05}. In the case of GPS radio galaxies, however, the UV disk emission toward the line of sight is likely to be absorbed by obscuring dusty tori.

Note that, in the framework of a model ascribing the spectral turnover of GPS sources to free-free absorption by the engulfed NLR clouds, it is the UV disk emission which is required to photoionize the absorbing matter \citep{beg99}. In addition to this, the UV disk radiation provides an important source of seed photons with energies $h \nu_0 \approx 10$\,eV for the inverse-Compton (IC) emission of ultrarelativistic lobe electrons. The volume-averaged energy density of this photon field, as a function of the distance from the central engine (and hence the source linear size $LS$), can be simply estimated as
\begin{eqnarray}  
U_{\rm UV} & = & {1 \over V} \, \int {L_{\rm UV} \over 4 \pi \, LS^2 \, c} dV = {L_{\rm UV} \over 2 \pi \, LS^2 \, c} \, \ln\left({LS_{\rm max} \over LS_{\rm min}}\right) \approx \nonumber \\
& \approx & 10^{-6} \, L_{\rm UV, \, 46} \, LS_{100}^{-2} \, {\rm erg \, cm^{-3}} \, ,
\end{eqnarray}
\noindent
where we put $\ln\left(LS_{\rm max}/LS_{\rm min}\right) \approx 3$ and $L_{\rm UV, \, 46} \equiv L_{\rm UV} / 10^{46}$\,erg\,s$^{-1}$. This is comparable to or slightly lower than the magnetic field energy density within the lobes, $U_{\rm B} \approx \eta_{\rm B} \, 10^{-6} \, L_{\rm j, \, 45}^{1/2} \, LS_{100}^{-1}$\,erg\,cm$^{-3}$ (see equation 2), for the expected values $L_{\rm j, \, 45} \geq 0.1$ and $L_{\rm UV, \, 46} \leq 10$.

Some part of the strong direct disk emission in powerful AGNs --- typically more than $10\%$ --- is expected to be reprocessed by the obscuring matter (dusty tori) and re-emitted at FIR-to-NIR frequencies. Several observations confirm the presence of such a spectral feature in the radiative outputs of young radio galaxies. For example, \citet{hec94} showed that GPS/CSS sources have the same MFIR strengths as extended sources with comparable radio powers and redshifts, i.e., $\langle L_{\rm 50 \, \mu m} \rangle \sim 3 \times 10^{45}$\,erg\,s$^{-1}$ for $\langle L_{\rm 5 \, GHz} \rangle \sim 10^{44}$\,erg\,s$^{-1}$ in the case of GPS/CSS radio galaxies \citep[see also][]{hes95}. We note, however, that the extended sources probably have lower efficiency in converting the jet kinetic power to synchrotron radio emission. \citet{fan00} confirmed that the dust emission dominating the radiative output at FIR frequencies is similar in the GPS/CSS population and regular radio galaxies, both in luminosity and temperature. This implies that the young radio sources do not contain more (or less) dust than the extended ones: on average, the FIR luminosities of these objects are $\langle L_{\rm FIR} \rangle \sim 3 \times 10^{44}$\,erg\,s$^{-1}$ and can be modeled by a two-temperature dust distribution with a total mass of $\sim 10^8 \, M_{\odot}$. These findings were recently confirmed by \citet{shi05}, who emphasized similarities in MFIR emission between CSS objects and regular radio-loud quasars or powerful radio galaxies.

The energy density of the photon field due to the dusty torus at a distance $r$ from the galactic center can be estimated as \citep[see][]{sik02,bla04}
\begin{equation}
U_{\rm IR}(r) = {L_{\rm IR} \over 4 \pi \, r_{\rm d}^2 \, c} \, { 1 \over 1 + (r / r_{\rm d})^2} \, ,
\end{equation}
\noindent
where $L_{\rm IR} \sim L_{\rm UV}$ is the expected torus luminosity, and $r_{\rm d} = (L_{\rm UV} / 4 \pi \, \sigma_{\rm SB} \, T_{\rm d}^4)^{1/2}$ is the characteristic (minimum) distance of the circumnuclear dust with temperature $T_{\rm d} \sim 10^3$\,K. Since $r_{\rm d}$ is supposed to be much smaller than the spatial scales considered here, $r_{\rm d} \approx L_{\rm UV, \, 46}^{1/2}$\,pc, one can restrict the analysis to $LS > r_{\rm d}$, thus obtaining the volume-averaged
\begin{equation}
U_{\rm IR} \approx 10^{-6} \, L_{\rm IR, \, 46} \, LS_{100}^{-2} \, {\rm erg \, cm^{-3}} \, ,
\end{equation}
\noindent
with $L_{\rm IR, \, 46} \equiv L_{\rm IR} / 10^{46}$\,erg\,s$^{-1}$. The mean value of the IR photon energy is hereafter assumed as $\nu_0 \approx 10^{13}$\,Hz. That follows from the fact that the dust temperature is $\propto r^{-1/2}$, while the dust density is inversely proportional to some high power of $r$. Therefore, the IC emission taking place at larger ($r > r_{\rm d}$) distances, as considered here, is mainly related to the hotter dust, i.e., to the MIR frequencies of the target photons.

The photon fields $U_{\rm UV}$ and $U_{\rm IR}$ evaluated above scale linearly with the nuclear luminosity $L_{\rm UV}$. So far, we have assumed a relatively high value of $L_{\rm UV} = 10^{46}$\,erg\,s$^{-1}$, as is appropriate for the nuclear spectra of quasar sources. It is not clear, however, whether the nuclei of \emph{all} GPS/CSS radio galaxies are indeed similar to the quasar ones. The emerging consensus is that powerful radio galaxies of FR~II type --- those possessing high-ionization broad and/or narrow lines --- are in fact misaligned quasars, with the UV accretion-related emission obscured toward the line of sight by dusty tori \citep{bar89}. This is supported by the fact that the obscured nuclear UV component is often observed in these objects indirectly via polarized scattered light or via intense re-emission of the obscuring matter at MFIR. In the framework of such a unification scheme, radio-loud quasars are eventually brighter in IR when compared to powerful FR~IIs only due to the contribution of the beamed jet emission \citep{haa04,shi05,ogl06,cle07}. Thus, powerful GPS/CSS sources which are precursors of powerful FR~II radio galaxies are not expected to lack the strong accretion-related UV and IR emissions discussed above. The situation is, however, less obvious in the case of young radio sources that are precursors of FR~I radio galaxies or of weaker FR~II objects with only low-ionization emission lines.

The radiative output of active nuclei in FR~I radio galaxies at optical frequencies is most likely dominated by the synchrotron emission of unresolved jets \citep{chi00}. They are generally believed to lack strong accretion-related optical/UV radiation \citep[but see][]{mao07}, in agreement with the weak nuclear obscuration advocated for these sources \citep{chi02}. Nuclei of lower-power FR~II sources are very often similar to nuclei of FR~I radio galaxies, with only low-ionization emission lines present, optical flux most likely dominated by the jet emission, and relatively moderate MFIR emission \citep{chi00,chi02,ogl06}. Thus, nuclear obscuration in these objects is believed to be only moderate \citep[although still present; see][]{haa05,har06}; in other words, accretion-related emission is believed to be lower than in powerful FR~IIs/quasars. Clearly, in the case of GPS/CSS radio galaxies that are precursors of these objects, kpc-scale photon fields might be lower than our model parameters $L_{\rm UV} \sim L_{\rm IR} \sim 10^{46}$\,erg\,s$^{-1}$. Even in this case, however, the synchrotron emission of the radio lobes provides a relatively intense photon field for the IC process. Its energy density can be evaluated as
\begin{equation}
U_{\rm syn} = {\int L_{\nu,\, {\rm syn}} \, d \nu \over 4 \pi \, l_{\rm c} \, LS \, c} \, ,
\end{equation}
\noindent
where the synchrotron luminosity (equation 10) is integrated over the whole frequency range with absorption effects included. Taking approximately $\int L_{\nu,\, {\rm syn}} \, d \nu \sim L_{\rm syn}$ as given by equation 9 (i.e., ignoring absorption of the radio photons), as well as $f\!(\gamma) \sim \gamma_{\rm cr}$ (see section \S\,3.1), one obtains
\begin{equation}
U_{\rm syn} \sim 5.2 \times 10^{-9} \, \eta_{\rm e} \, L_{\rm j, \, 45}^{3/4} \, LS_{100}^{-3/2} \quad {\rm erg \, cm^{-3}} \, ,
\end{equation}
\noindent
which is again less than or, at most, comparable to the magnetic field energy density $U_{\rm B}$.

Yet another source of seed photons for IC scattering within the lobes of young radio objects is provided by the optical light of host galaxies. In general, the host galaxies of GPS/CSS sources are very similar to the hosts of powerful (3CR) classical doubles when observed in NIR \citep{dev98,dev00}, being evolved ellipticals with some morphological indications of relatively recent merger events. Very often, they also exhibit kpc-scale optical emission aligned with the main axis of the radio source. \citet{axo00} and \citet{lab05} found that the aligned optical features consist of at least three distinct components, namely an unresolved nucleus, line-emission aligned with and of similar extent as the radio lobes, and a weak component extending beyond the radio lobes. The unresolved optical nuclei are present only in the sources with broad emission lines, in agreement with the nuclear obscuration/unification scheme. The kpc-scale line-emitting gas is most probably due to the interaction of gas clouds with the expanding jets and lobes. Finally, the diffuse component extending beyond the radio lobes is most likely photoionized by anisotropic nuclear emission. Hereafter, we consider conservatively strictly the starlight emission of the hosts, which corresponds to NIR frequencies of $\nu_0 \approx 10^{14}$\,Hz. For a given $V$-band luminosity of the galaxy, $L_{\rm V}$, which is known to correlate well with the core radius of the stellar distribution, $r_{\rm s}$, namely $L_{\rm V}/10^{45}$\,erg\,s$^{-1}$$ \approx r_{\rm s}/1$\,kpc \citep{der05}, the starlight energy density may be evaluated roughly as
\begin{equation}
U_{\rm star} = {3 \, L_{\rm V} \over 4 \pi \, r_{\rm s}^2 \, c} \approx 8 \times 10^{-10} \, L_{\rm V, \, 45}^{-1} \, {\rm erg \, cm^{-3}} \, ,
\end{equation}
\noindent
where $L_{\rm V, \, 45} \equiv L_{\rm V}/10^{45}$\,erg\,s$^{-1}$, and we assumed approximately $LS \leq r_{\rm s}$.

\subsection{Inverse-Compton Emission}

In the Thomson regime of the IC scattering, the IC luminosities related to the external photon fields can be simply evaluated as
\begin{equation}
[\nu L_{\nu}]_{\rm IC/rad} = {2 \over 3} c \, \sigma_T \, V \, U_{\rm rad} \, \left[ \gamma^3 \, N_{\rm e}(\gamma) 
\right]_{\gamma = \sqrt{3 \nu / 4 \nu_0}} \, .
\end{equation}
\noindent
Here $U_{\rm rad}$ stands for the energy densities of different seed radiation sources ($U_{\rm UV}$, $U_{\rm IR}$, or $U_{\rm star}$), which are approximated as being monochromatic with characteristic seed photon frequencies $\nu_0$ as given in \S\,3.2. In the case of SSC emission, the monochromatic approximation is not correct, and the appropriate luminosity can be evaluated as
\begin{equation}  
[\nu L_{\nu}]_{\rm ssc} \approx {2 \over 3} \, \sigma_{\rm T} \, {l_{\rm c} \over 3} \, \int_{3 \nu / 4 \gamma_{\rm max}^2}^{3 \nu / 4 \gamma_{\rm min}^2} \,  L_{\nu_0, \, {\rm syn}} \, \left[ \gamma^3 \, N_{\rm e}(\gamma) \right]_{\gamma = \sqrt{3 \nu / 4 \nu_0}} \, d\nu_0 \quad .
\end{equation}
\noindent
The resulting broad-band spectra are shown in Figures~2--3 for $L_{\rm V} = 10^{45}$\,erg\,s$^{-1}$, different jet kinetic powers $L_{\rm j}$ (i.e., $10^{47}$, $10^{46}$, $10^{45}$, and $10^{44}$\,erg\,s$^{-1}$), different source linear sizes $LS$ (i.e., $33$\,pc, $100$\,pc, and $1$\,kpc), and fixed lobe parameters $\eta_{\rm B} = 0.3$, $\eta_{\rm e} = 3$. As illustrative values we also take $L_{\rm UV} = L_{\rm IR} = 10^{46}$\,erg\,s$^{-1}$ for $L_{\rm j} > 10^{45}$\,erg\,s$^{-1}$, while $L_{\rm UV} = L_{\rm IR} = 10^{45}$\,erg\,s$^{-1}$ for $L_{\rm j} \leq 10^{45}$\,erg\,s$^{-1}$, accordingly to the discussion in section \S\,3.2. Different figures correspond to the different source functions $Q(\gamma)$, namely, to the single power-law injection with spectral index $s = 2.5$ (Figure~2) or to the broken-power-law with $s_1 = 1.5$, $s_2 = 3$ and the break energy $\gamma_{\rm int} = m_{\rm p}/m_{\rm e}$ (Figure~3). The absorption, radiative cooling, and adiabatic cooling effects in the electron energy distribution $N_{\rm e}(\gamma)$ are taken into account as described in \S\,2.3 and \S\,3.1. Finally, we assume minimum and maximum electron Lorentz factors $\gamma_{\rm min} = 1$ and $\gamma_{\rm max} = 10^5$, respectively. Solid lines in the figures correspond to the synchrotron component, dashed lines to the SSC emission, dotted lines to comptonization of the IR torus radiation, dot-dashed lines to comptonization of the starlight photon field, and short-dashed lines to comptonization of the UV disk emission.

Although the model predictions regarding the high-energy IC radiation may differ from source to source due to the scatter in the model parameters, one can outline some of the main results of the analysis. First, the evaluated SSC, IC/IR and IC/UV components are together very pronounced at $\sim$\,keV photon energies. One should note, at this point, that the emission from accretion disks in AGNs extends as well to X-ray photon energies (being produced by Comptonization of thermal optical/UV disk radiation within the hot disk's corona) and that this emission may be strong enough to dominate the inverse-Compton X-ray emission analyzed in this paper. \citet{kor99} showed that, in the case of quasar sources, the $1$\,keV disk luminosity is on average $\sim 10^{1.5}$ times lower than the UV-bump luminosity. This, for $L_{\rm UV} = 10^{46}$\,erg\,s$^{-1}$, is still high enough to dominate the X-ray lobes' emission evaluated above in all but the most powerful and compact GPS quasars. In addition, the contribution from unresolved relativistic jets to the total X-ray output may be also non-negligible in these objects. Therefore, the keV inverse-Compton emission from kpc-scale lobes is expected to be pronounced in powerful GPS radio galaxies rather than GPS quasars since, in the former, the direct X-ray emission is expected to be absorbed by the obscuring tori and the jet contribution is expected to be Doppler-hidden. In a subsequent paper, we investigate in more detail whether the IC lobe emission can indeed account for most of the X-ray fluxes detected from GPS radio galaxies, with the intrinsic, i.e. absorption-corrected, X-ray luminosities typically in the range $L_{\rm X} \approx 10^{42}-10^{45}$\,erg\,s$^{-1}$ and the fitted X-ray spectral indices $0 \lesssim \alpha_{\rm X} \lesssim 1.5$.  Interestingly, the lower-energy tails of the SSC, IC/IR and IC/UV spectral components may account for (or at least contribute significantly to) the extended UV emission detected in a number of GPS/CSS radio galaxies, with observed luminosities $L_{\rm X} \approx 10^{40}-10^{42}$\,erg\,s$^{-1}$ \citep{lab08}.

We note, in this context, that the initial low detection rate of GPS/CSS quasars at hard X-ray frequencies suggested that these objects were either intrinsically X-ray weak or heavily obscured. \citet{odea00} and \citet{gua04} argued for the nuclear (disk) origin of the observed keV photons and significant internal obscuration, with the required absorbing column densities being as large as $N_{\rm H} \sim 10^{24}$\,cm$^{-2}$ in some cases. Complex X-ray spectra complicate the interpretation of typically weakly detected spectral features \citep[see][for the case of Mkn 668]{gua04}. The most recent X-ray observations seem to confirm that GPS/CSS sources are obscured rather than intrinsically weak in X-rays \citep{gua06,vin06}. The implied obscuration is consistent with the presence of nuclear tori, with inferred absorbing column densities ranging from $N_{\rm H} \gtrsim 10^{20}$\,cm$^{-2}$ up to $N_{\rm H} \lesssim 10^{24}$\,cm$^{-2}$, but only in a few cases. Such obscuration is similar to that observed in extended and powerful radio galaxies, suggesting a very similar nuclear environment in GPS and FR\,II objects. As noted by \citet{vin06}, in all of the cases where the appropriate comparison could be performed, the evaluated $N_{\rm H}$ values exceed substantially (by a factor of $>10$) the absorbing column densities $N_{{\rm H{\sc I}} }$ implied by the detected H{\sc i} absorption lines. Such behavior may, in fact, be expected, taking into account the different locations of the absorbing media involved (nuclear tori vs. NLR clouds) and the different states of the absorbing matter (total vs. neutral fraction of the gas). 

Interestingly, in some CSS objects, clear excess (flat power-law) emission was noted at hard X-rays and ascribed to emission from the inner parts of the radio outflow. This is the case in quasar 3C 48 \citep{wor04}, Seyfert 1 object PKS 2004-447 \citep{gal06}, and radio galaxy 3C 303.1 \citep{odea06}. We note that the first two sources are especially bright in the infrared, most probably due to emission from the obscuring torus, and that this radiation can provide the dominant seed photon population for efficient X-ray inverse-Compton emission in the inner (pc-scale) portion of relativistic jets in \emph{quasar} sources \citep[see in this context][]{gua04,gal06}. Such a scenario was, in fact, proposed (and successfully applied) for the particular case of GPS quasar PKS 1127-145 \citep{sie02,bla04}.

Another interesting finding of our analysis is that the inverse-Comptonization of the UV disk emission seems to be strong enough to be detectable by {\it GLAST} for many GPS radio galaxies.\footnote{At lower $\gamma$-ray energies ($< 10$\,MeV), however, the IC emission from lobes discussed here may be overwhelmed by the `cocoon bremsstrahlung' radiation discussed by \citet{kin07}.} Note that the estimates presented in this paper are rather conservative because we have assumed roughly equal energy stored in the lobe magnetic field and in ultrarelativistic electrons, with the magnetic field intensity being a factor of only a few below the equipartition value \citep[$\eta_{\rm B} = 0.3$, $\eta_{\rm e} = 3$, in analogy to the lobes of classical doubles][and references therein]{kat05,cro05}. Any \emph{larger} deviations from the minimum power condition $\eta_{\rm B} \ll \eta_{\rm e}$ would increase the expected inverse-Compton emission for a given synchrotron power. Thus, one can conclude that young precursors of powerful radio sources will most likely constitute a very numerous class of extragalactic $\gamma$-ray sources that will be detected by {\it GLAST}. It is interesting to mention, in this context, that in the most recent sample of radio-loud quasars extracted from a combination of the SDSS (DR3) and FIRST surveys by \citet{dev06}, the majority of objects are characterized by a compact radio morphology. As pointed out by the authors, beaming effects on a parent population of extended sources cannot account for the observed compact-source counts. Thus, many of these objects have to be truly compact, and a fraction of them must be young and powerful GPS sources. Again, in the case of such compact GPS quasars, the IC emission from the unresolved jets may dominate the lobes' radiative output at GeV photon energies \citep[see][]{bla04,bai05}.

In order to quantify prospects for the detection of GPS radio galaxies in $\gamma$-rays, let us estimate roughly the high-energy segment of the expected IC/UV emission, which dominates the lobes' radiative output at these photon energies. Assuming a single power-law injection with relatively steep electron energy index $s=2.5$, corresponding to the `radiatively cooled' emission spectral index $\alpha = {1 \over 2} \, (s-1) +0.5 = 1.25$, one may find from equations 17, 19, and 25 the monochromatic IC/UV luminosity
\begin{equation}  
{[\varepsilon L_{\varepsilon}]_{\rm IC/UV} \over 10^{42} \, {\rm erg/s}} \sim 2 \, \left({\eta_{\rm e} \over \eta_{\rm B}}\right) \left({L_{\rm j} \over 10^{45} \, {\rm erg/s}}\right)^{1/2} \left({LS \over 100 \, {\rm pc}}\right)^{-1} \left({L_{\rm UV} \over 10^{46} \, {\rm erg/s}}\right) \left({\varepsilon \over 1 \, {\rm GeV}}\right)^{-0.25}
\end{equation}
\noindent
or the appropriate IC/UV energy flux
\begin{equation}
{[\varepsilon S_{\varepsilon}]_{\rm IC/UV} \over 10^{-12} \, {\rm erg / cm^{2} / s}} \sim 1.6 \times \left({[\varepsilon L_{\varepsilon}]_{\rm IC/UV} \over 10^{42} \, {\rm erg/s}}\right) \left({d_{\rm L} \over 100 \, {\rm Mpc}}\right)^{-2} ,
\end{equation}
\noindent
where $d_{\rm L}$ is the luminosity distance to the source. This can be compared with sensitivities of modern and planned $\gamma$-ray instruments. For example, \citet{pan07} gives the differential sensitivity of the {\it GLAST} LAT instrument, defined as `the flux level over an energy interval of 1/4 of a decade over which the statistical significance is 2 standard deviations in 1 year' roughly $\sim 0.8 \times 10^{-12}$\,erg\,cm$^{-2}$\,s$^{-1}$. This implies that GPS radio galaxies can be detected at $1$\,GeV photon energies by {\it GLAST} in its 1-year all-sky survey if $d_{\rm L} \leq 100 \, \left(\eta_{\rm e} / \eta_{\rm B}\right)^{1/2} L_{\rm j, \, 45}^{1/4} \, LS_{100}^{-1/2} \, L_{\rm UV, \, 46}^{1/2}$\,Mpc. Thus, the most compact ($LS < 100$\,pc) and powerful ($L_{\rm j} > 10^{46}$\,erg\,s$^{-1}$) sources with energy equipartition in their lobes fulfilled ($\eta_{\rm e} \gtrsim \eta_{\rm B}$) are expected to be visible at GeV photon energies up to distances of $\sim 1$\,Gpc. Note that even if the accretion disc luminosities in these objects are low, $L_{\rm UV, \, 46} < 1$, the SSC component may provide a very strong source of target UV photons for the IC scattering. We also mention that since the sensitivities of EGRET and {\it AGILE} instruments are lower than the planned sensitivity of the {\it GLAST} LAT mission, one should not expect GPS radio galaxies to be detected by them. Obviously, there may be some exceptions, and it is possible that several unidentified high-latitude non- (or weakly) variable EGRET sources are associated with young radio objects. We emphasize that the estimates 27$-$28 provided above are very conservative, and any larger deviation from the energy equipartition, $\eta_{\rm e} \gg \eta_{\rm B}$, as well as some other spectral shape of the electron injection function (e.g., the broken power-law discussed in the previous sections), will substantially increase the expected IC/UV flux. Interestingly, if the IC/UV component as discussed above (equation 28) extends to even higher-energy $\gamma$-rays, the expected IC/UV fluxes of GPS radio galaxies are accessible by the modern ground-based Cherenkov telescopes. For example, a rough detection limit for H.E.S.S. or MAGIC with a reasonable on-source exposure is $\sim 10^{-13}$\,erg\,cm$^{-2}$\,s$^{-1}$ at 1\,TeV photon energies, implying that GPS radio galaxies could, in principle, be detected by these instruments for source distances, again, $d_{\rm L} \leq 100 \, \left(\eta_{\rm e} / \eta_{\rm B}\right)^{1/2} L_{\rm j, \, 45}^{1/4} \, LS_{100}^{-1/2} \, L_{\rm UV, \, 46}^{1/2}$\,Mpc.

\section{Conclusions}

There is an emerging agreement that compact radio sources characterized by inverted low-frequency radio spectra and symmetric (`lobe-lobe') radio morphology are young rather than frustrated by an unusually dense ambient medium. That is because recent multiwavelength observations reveal repeatedly that the properties of their environments and of their central engines are very similar to the ones found in the extended radio sources. Here we utilize these newest multifrequency data, discussing evolution and non-thermal broad-band emission of the lobes in GHz-peaked-spectrum (GPS) radio galaxies. First, we propose a simple dynamical model for the GPS sources, exploiting the standard set of equations describing evolution of a relativistic jet---cocoon system. In contrast to the scenarios analyzed previously in the literature, however, we assume a uniform distribution for the ambient medium with number density $n_0 \approx 0.1$\,cm$^{-3}$, as observed in central parts ($<1$\,kpc) of elliptical hosts of radio galaxies. This gives a constant advance velocity of the jet, which we fix as $v_{\rm h} \approx 0.1\,c$, consistent with radio studies of the hotspots' proper motions in several GPS objects. With these few model assumptions we calculate all the other lobe parameters as functions solely of the jet kinetic power, $L_{\rm j}$, and the source linear size, $LS$. In particular, the evaluated equipartition lobe magnetic field intensity $B \sim 1$\,mG is consistent with the value implied by the spectral-ageing analysis of radio emission from GPS galaxies, and the implied sideways expansion velocity of the lobes, $v_{\rm c} \gtrsim 10^8$\,cm\,s$^{-1}$, agrees with the outflow velocities of the line-emitting gas observed in many young radio objects. We note that both $B$ and $v_{\rm c}$ depend weakly on the unknown jet power $L_{\rm j}$, namely $\propto L_{\rm j}^{1/4}$, which is a comfortable feature of the model.

In the framework of the proposed dynamical description of young and compact radio sources, we follow the evolution of ultrarelativistic electrons injected from a terminal jet hotspot to the expanding lobes, taking the appropriate adiabatic and radiative energy losses into account. We find that, for a variety of the injected electron spectral shapes (which may be, in fact, different from the often assumed single power law $\propto \gamma^{-2}$), the resulting electron energy distribution within the lobes is expected to be of a broken power-law form, with critical electron energy $\gamma_{\rm cr} \sim 200 \, \left(L_{\rm j} / 10^{45} \, {\rm erg \, s^{-1}}\right)^{-1/2}$. We argue that some plateau (or `non-power-law curvature') around GHz frequency range is to be expected in the synchrotron spectra of GPS sources as a result of such a break. We also find that the total (unabsorbed) synchrotron luminosity of the lobes is expected to be constant in time during the GPS phase of the evolution (or, in other words, is expected to be independent on the source linear size), constituting about $10\%$ (or more) of the total kinetic power of the jet. As for the formation of the inverted low-frequency radio spectra, we argue that it cannot be due to synchrotron self-absorption effects, because these cannot reproduce neither the slope nor the normalization of the observed $\nu_{\rm p} \propto LS^{-0.65}$ anticorrelation. Instead, we favor free-free absorption of radio photons by the neutral clouds of the interstellar medium, engulfed by the expanding lobes and photoionized by emission from the active center, as proposed previously by \citet{beg99}. We note that these clouds may be naturally identified with the ones producing narrow-line emission (NLR) and also H{\sc i} absorption lines observed in many GPS radio sources. We speculate that both the observed anticorrelation of the neutral column density (corresponding to the observed H{\sc i} absorption lines) with the source linear size, $N_{\rm H{\sc I}} \propto LS^{-0.45}$, as well as the anticorrelation of the spectral turnover frequency (being a result of a free-free absorption by the engulfed clouds) with the source linear size, is primarily due to the density decrease of the NLR clouds with the distance from the active center. Such a decrease, $\propto r^{-n}$ with $1<n<2$, is observed directly in several nearby Seyfert galaxies.

We estimate different photon fields within the lobes of GPS radio galaxies, due to emission from accretion disks, obscuring nuclear tori, synchrotron radiation of the lobe electrons, and starlight from the elliptical hosts. We calculate the resulting inverse-Compton components for different electron injection conditions, different jet powers, and different source linear sizes. We find complex high-energy spectra, extending from eV up to GeV (and possibly even TeV) photon energies. The resulting fluxes seem to be strong enough to dominate (or at least contribute significantly to) the radiation observed from GPS objects at UV and X-ray frequencies. We expect that in the case of GPS radio galaxies --- where, due to large inclinations, the direct disk emission in the UV/X-ray range is expected to be obscured toward the line of sight by the dusty tori, while any jet radiative contribution is expected to be Doppler-hidden --- the discussed non-thermal emission from the lobes is expected to be especially pronounced. In a subsequent paper, we investigate in more detail this issue through the analysis of the observed spectra of a sample of young and compact radio galaxies detected in the X-ray domain. Finally, we find that GPS radio galaxies are expected to be bright in the GeV photon energy range and that they can be detected by the forthcoming {\it GLAST} mission up to distances of the order of (conservatively) $\sim 1$\,Gpc. Since the population of young radio sources is numerous when compared to the population of the extended ones, and since the high-energy lobe emission discussed in this paper is isotropic, one can expect that GPS radio galaxies are likely to constitute a numerous class of extragalactic {\it GLAST} sources.

\acknowledgments
{\L}.S. acknowledges support by the MEiN grant 1-P03D-003-29. L.O. gratefully acknowledges partial support from the INFN grant PD51, and thanks the Department of General Physics of the University of Torino for hospitality during the 2007 pre-fellow period. R.M. acknowledges support from MNiSW grant no. 1-P03D-009-28. J.K. acknowledges support by JSPS KAKENHI (19204017/14GS0211). L.O. and S.W. acknowledge support by BMBF/DLR through grant 50OR0303. {\L}.S., L.O. and S.W. were partly supported by the ENIGMA Network through the grant HPRN-CT-2002-00321. The authors thank G.V. Bicknell, M. Sikora, S.E. Healey, and the anonymous referee for very helpful comments.

\appendix

\section{Electron Evolution}

Let us consider the kinetic equation
\begin{equation}
{\partial \mathcal{N}_{\rm e}(\gamma, t) \over \partial t} = {\partial \over \partial \gamma} \left\{\left|\dot{\gamma}\right| \, \mathcal{N}_{\rm e}(\gamma, t)\right\} + Q(\gamma, t)
\end{equation}
\noindent
describing the time evolution of the electron energy distribution $\mathcal{N}_{\rm e}(\gamma, t) = N_{\rm e}(\gamma, t) \, V$ under the influence of radiative (both synchrotron and inverse-Compton/Thomson-regime) as well as adiabatic energy losses,
\begin{equation}
\left|\dot{\gamma}\right| = {1 \over 3} {\dot{V} \over V} \, \gamma + c_1 \, U \, \gamma^2 \, ,
\end{equation}
\noindent
where $Q(\gamma, t)$ is the injection function, $V$ is the volume of the system, $\dot{V} = dV / dt$, $c_1 = 4 \, \sigma_{\rm T} / 3 \, m_{\rm e} c$, and $U = U_{\rm B} + U_{\rm rad}$ is the sum of the magnetic field and radiation field energy densities. Here we consider $V = V(t)$, $\dot{V} = \dot{V}(t)$, and $U=U(t)$.

Using the method of characteristics, one can find the solution to the kinetic equation A1 in a form
\begin{equation}
\mathcal{N}_{\rm e}(\gamma, t) = e^{\int^t \left({1 \over 3} {\dot{V} \over V} + 2 \, c_1 \, \gamma \, U \right) \, dt'} 
\times \left\{\mathcal{N}_{\rm e}\left(\gamma_0, t=0\right) + \int^t Q[\gamma(t'), t')] \,\,\, 
e^{- \int^{t'} \left({1 \over 3} {\dot{V} \over V} + 2 \, c_1 \, \gamma \, U \right) \, dt''} \, dt' \right\} \, ,
\end{equation}
\noindent
where
\begin{equation}
\gamma(t) = {\exp\left[ - \int^t {1 \over 3} {\dot{V} \over V} \, dt' \right] \over \gamma_0^{-1} + c_1 \, \int^t
\exp\left[ - \int^{t'} {1 \over 3} {\dot{V} \over V} \, dt'' \right] \, U \, dt'} \, ,
\end{equation}
\noindent
and $\gamma_0 \equiv \gamma(t=0)$. Since 
\begin{equation}
e^{\int^t {1 \over 3} {\dot{V} \over V} \, dt'} = \left({V \over V_0}\right)^{1/3} \, ,
\end{equation}
\noindent
where $V_0 \equiv V(t=0)$, one can find that A4 reads as
\begin{equation}
\gamma(t) = {\gamma_0 \, \left({V \over V_0}\right)^{-1/3} \over 1 + c_1 \, \gamma_0  \int^t \, U \, \left({V \over V_0}\right)^{-1/3} dt'} \, .
\end{equation}
\noindent
Similarily, noting that
\begin{equation}
e^{2\, c_1 \int^t \gamma \, U \, dt'} = \left[1 + c_1 \, \gamma_0 \, \int^t \, U \, \left({V \over V_0}\right)^{-1/3} dt' \right]^2 \, .
\end{equation}
\noindent
one can rewrite A3 in the simpler form
\begin{equation}
\mathcal{N}_{\rm e}(\gamma, t) = \gamma^{-2} \, \left({V \over V_0}\right)^{-1/3} \int^t Q[\gamma(t'), t'] \,\,\, \gamma(t')^2 \, \left({V \over V_0}\right)^{1/3} dt' \, ,
\end{equation}
\noindent
where we set $\mathcal{N}_{\rm e}\left(\gamma_0, t=0\right) = 0$. Note that in the above integral one has to substitute $\gamma(t')$ with the expression A6, and then, after integrating, replace back $\gamma_0$ with $\gamma$.

\clearpage

\begin{figure*}
\center{
\includegraphics[scale=0.5,angle=270]{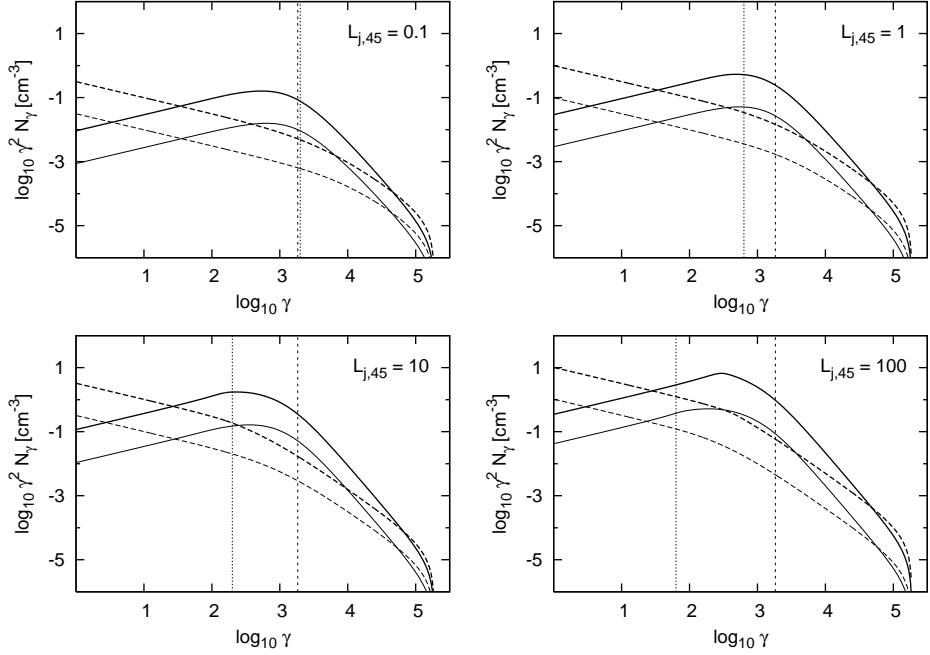}
}
\figcaption{
Evolution of the electron energy distribution within the lobes of GPS sources for different jet powers ($L_{\rm j} = 10^{44}$, $10^{45}$, $10^{46}$, and $10^{47}$\,erg s$^{-1}$) and different source linear sizes ($LS = 100$\,pc and $1$\,kpc; thick/upper and thin/lower lines, respectively). Two different source functions were considered for illustration, namely a single power-law $Q(\gamma) \propto \gamma^{-2.5}$ (dotted lines) or a broken power-law $Q(\gamma) \propto \gamma^{-1.5}$ for $\gamma < \gamma_{\rm int}$ and $Q(\gamma) \propto \gamma^{-3}$ for $\gamma > \gamma_{\rm int}$ (solid lines). In both cases, the minimum and maximum electron Lorentz factors are $\gamma_{\rm min} = 1$ and $\gamma_{\rm max} = 10^5$ while the normalization of the injection function is evaluated through the condition $U_{\rm e} = \eta_{\rm e} \, p$ with $\eta_{\rm e} = 3$ and $\eta_{\rm B} = 0.3$. Vertical dotted and dashed lines indicate critical electron energies $\gamma_{\rm cr} = 200 \, \eta_{\rm B}^{-1} \, L_{\rm j, \, 45}^{-1/2}$ and $\gamma_{\rm int} = m_{\rm p} / m_{\rm e}$, respectively.
} 
\end{figure*}

\clearpage
\pagestyle{empty}
\setlength{\voffset}{-20mm}
\begin{figure*}
\center{
\includegraphics[scale=0.4]{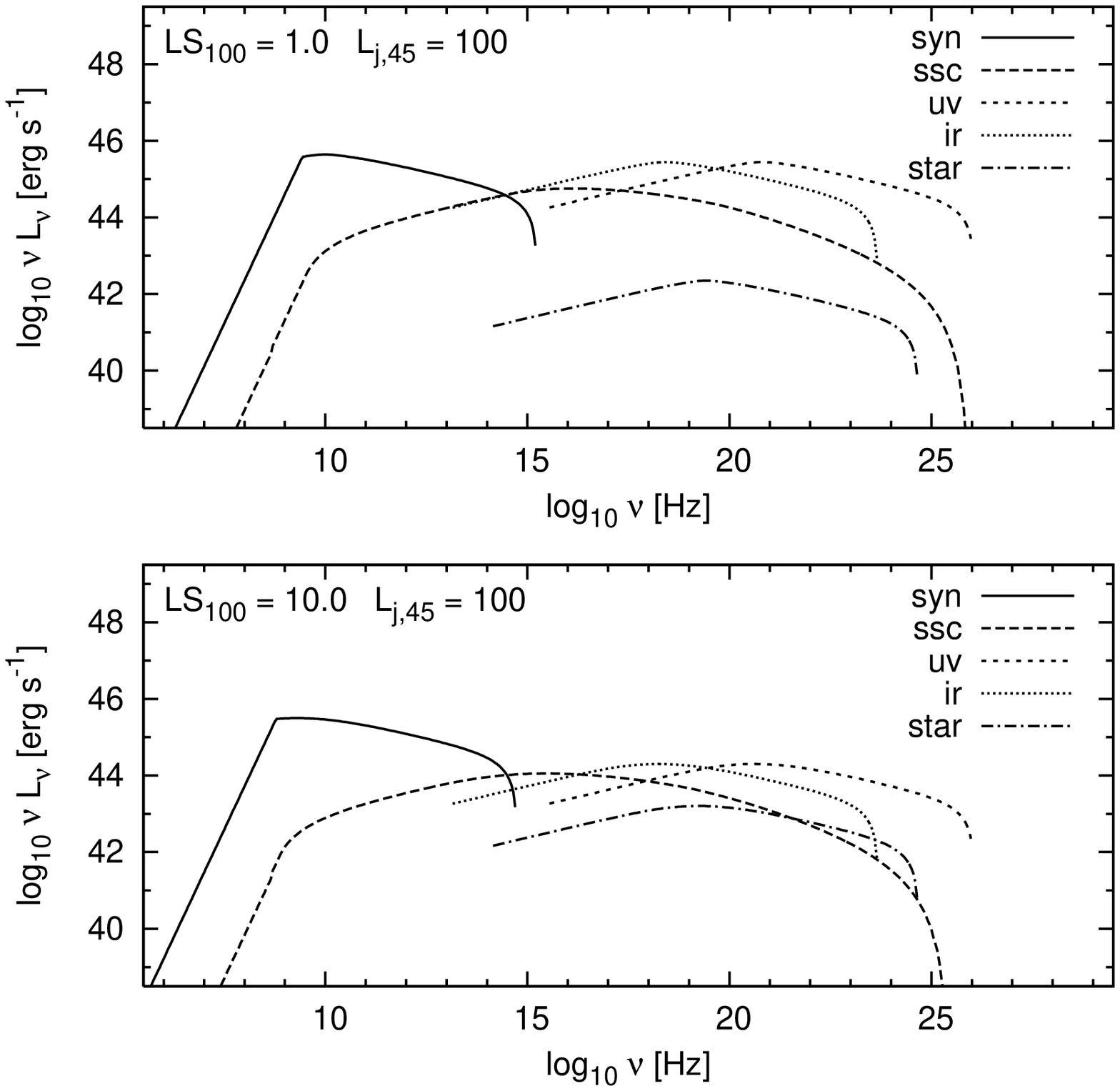}
\includegraphics[scale=0.4]{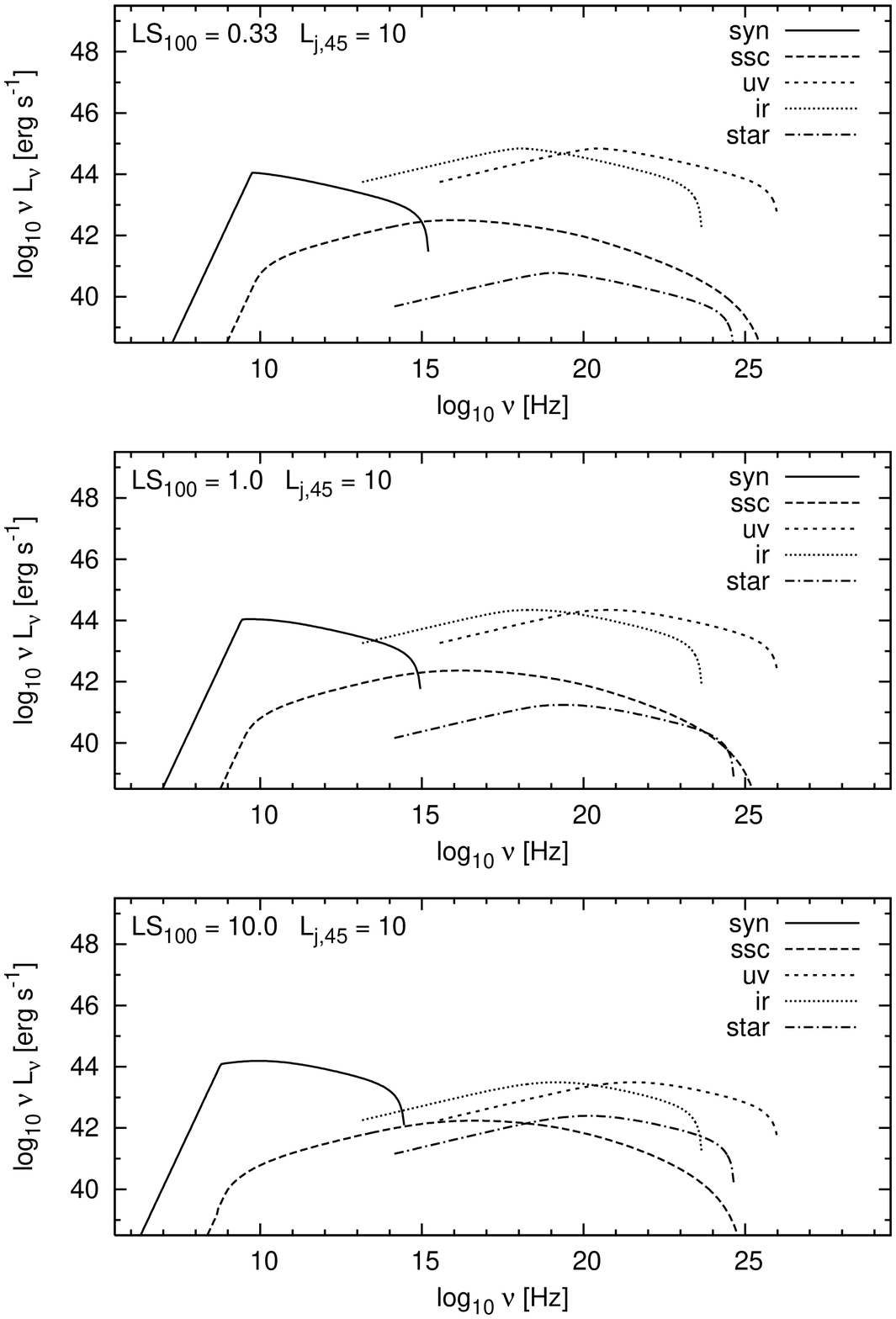}
\includegraphics[scale=0.4]{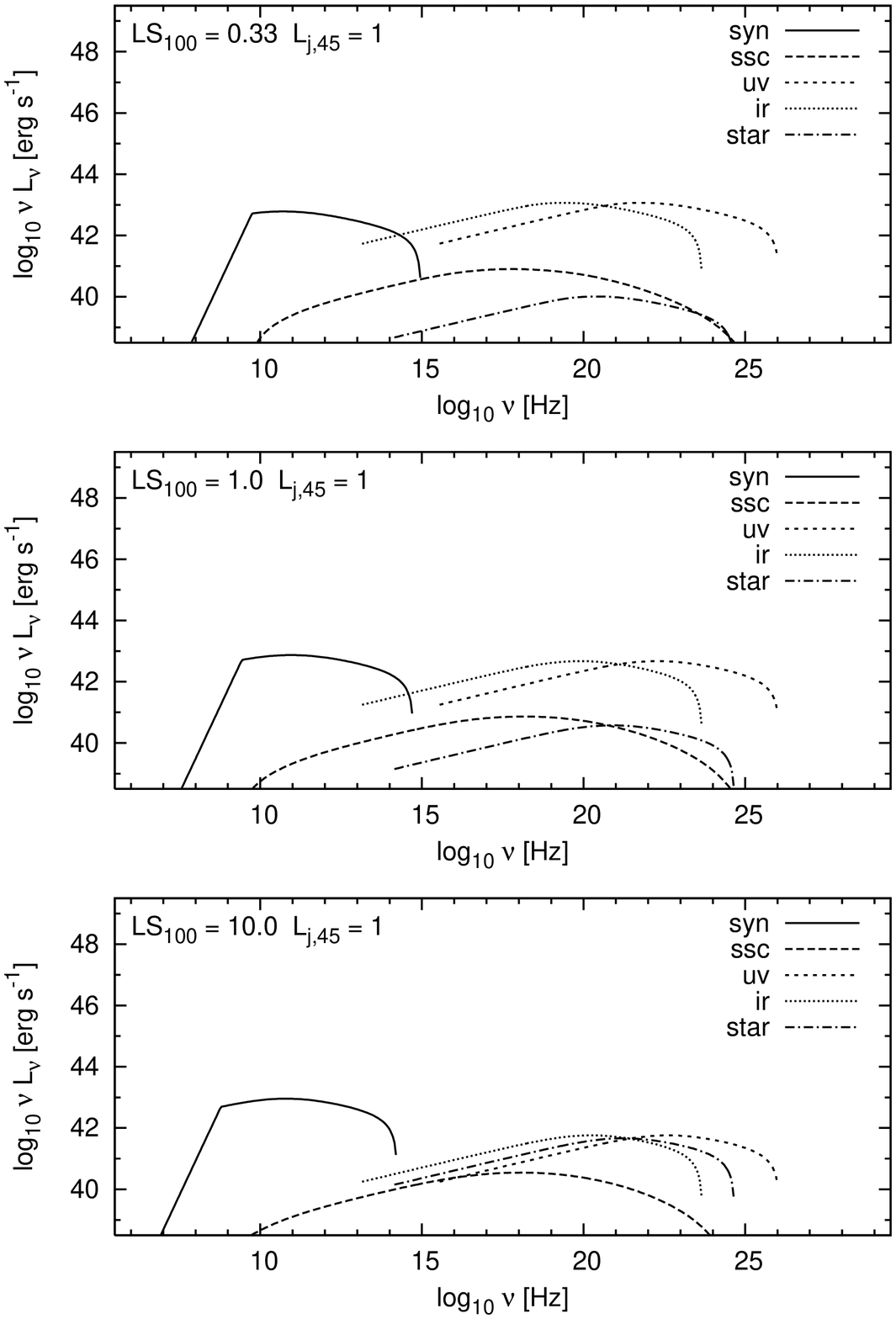}
\includegraphics[scale=0.4]{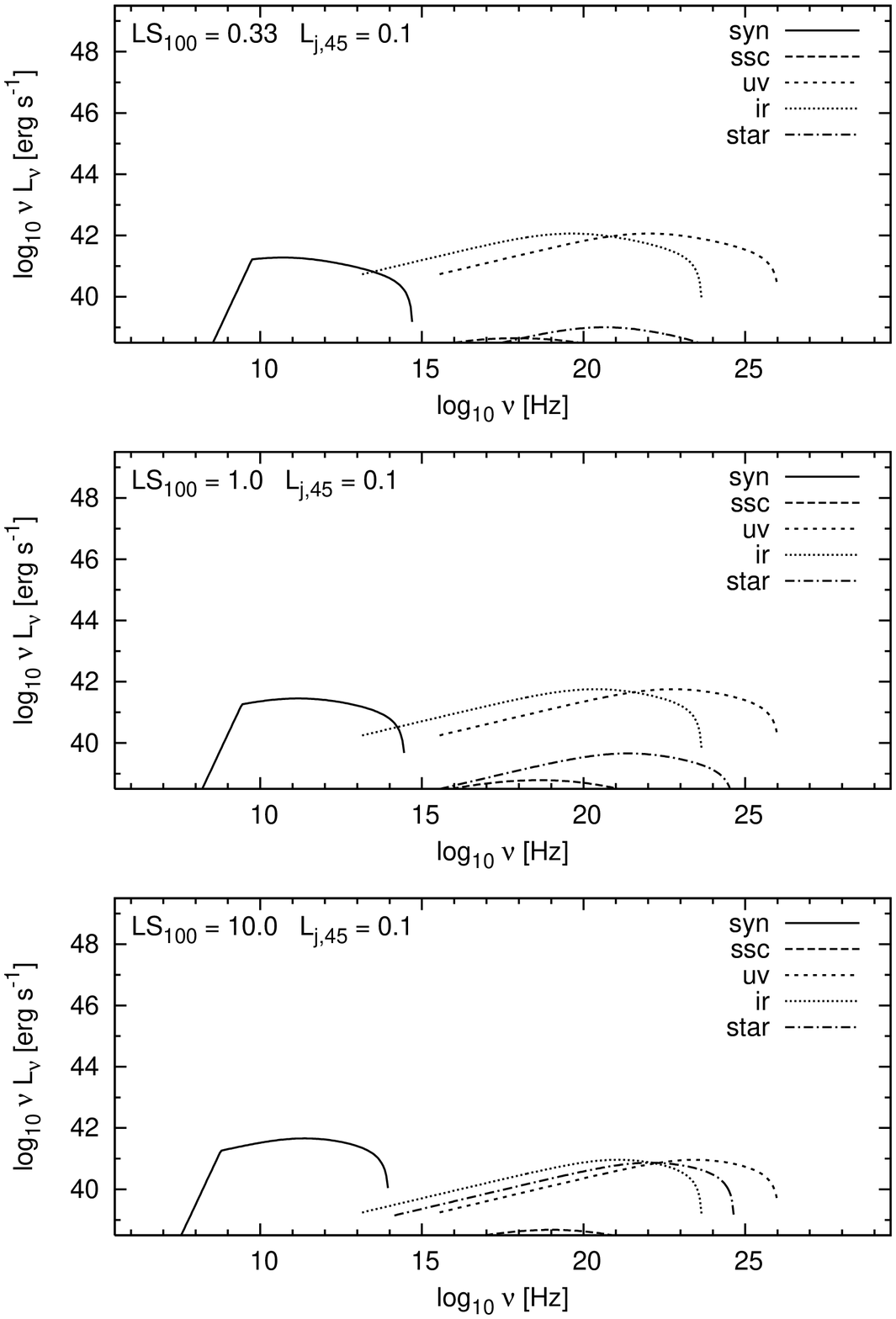}
}
\figcaption{
Broad-band emission produced within the lobes of GPS sources with different jet kinetic power ($L_{\rm j} = 10^{47}$\,erg\,s$^{-1}$, $10^{46}$\,erg\,s$^{-1}$, $10^{45}$\,erg\,s$^{-1}$, $10^{44}$\,erg\,s$^{-1}$) and different linear sizes ($LS=33$\,pc, $100$\,pc, and $1$\,kpc). Illustrative parameters were considered: $\eta_{\rm B} = 0.3$, $\eta_{\rm e} = 3$, $L_{\rm V} = 10^{45}$\,erg\,s$^{-1}$, $L_{\rm UV} = L_{\rm IR} = 10^{46}$\,erg\,s$^{-1}$ for $L_{\rm j} > 10^{45}$\,erg\,s$^{-1}$, and $L_{\rm UV} = L_{\rm IR} = 10^{45}$\,erg\,s$^{-1}$ for $L_{\rm j} \leq 10^{45}$\,erg\,s$^{-1}$. Single power-law injection function $Q(\gamma)$ with spectral index $s = 2.5$ was assumed, with minimum and maximum electron Lorentz factors $\gamma_{\rm min} = 1$ and $\gamma_{\rm max} = 10^5$, respectively.
} 
\end{figure*}

\begin{figure*}
\center{
\includegraphics[scale=0.4]{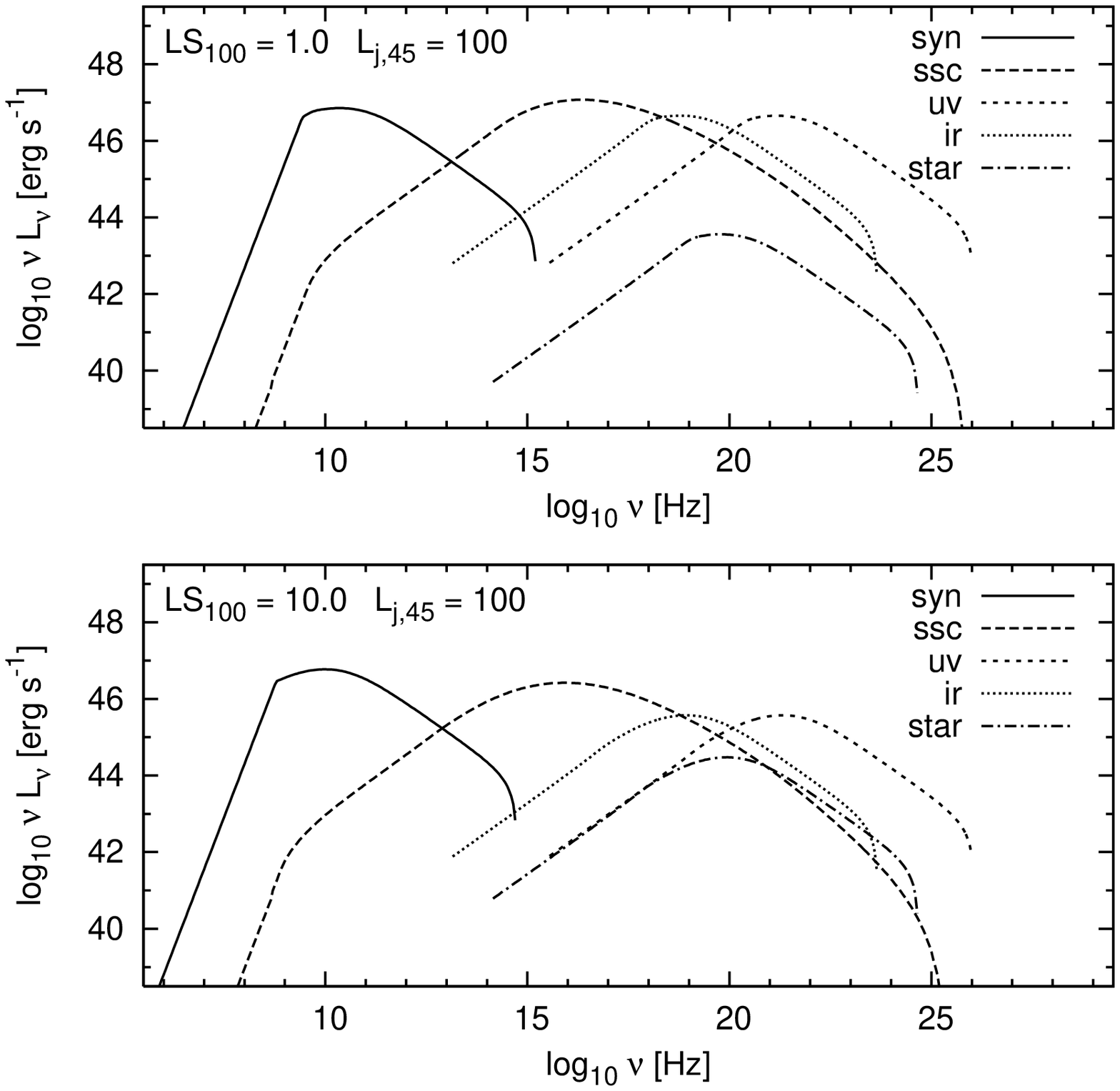}
\includegraphics[scale=0.4]{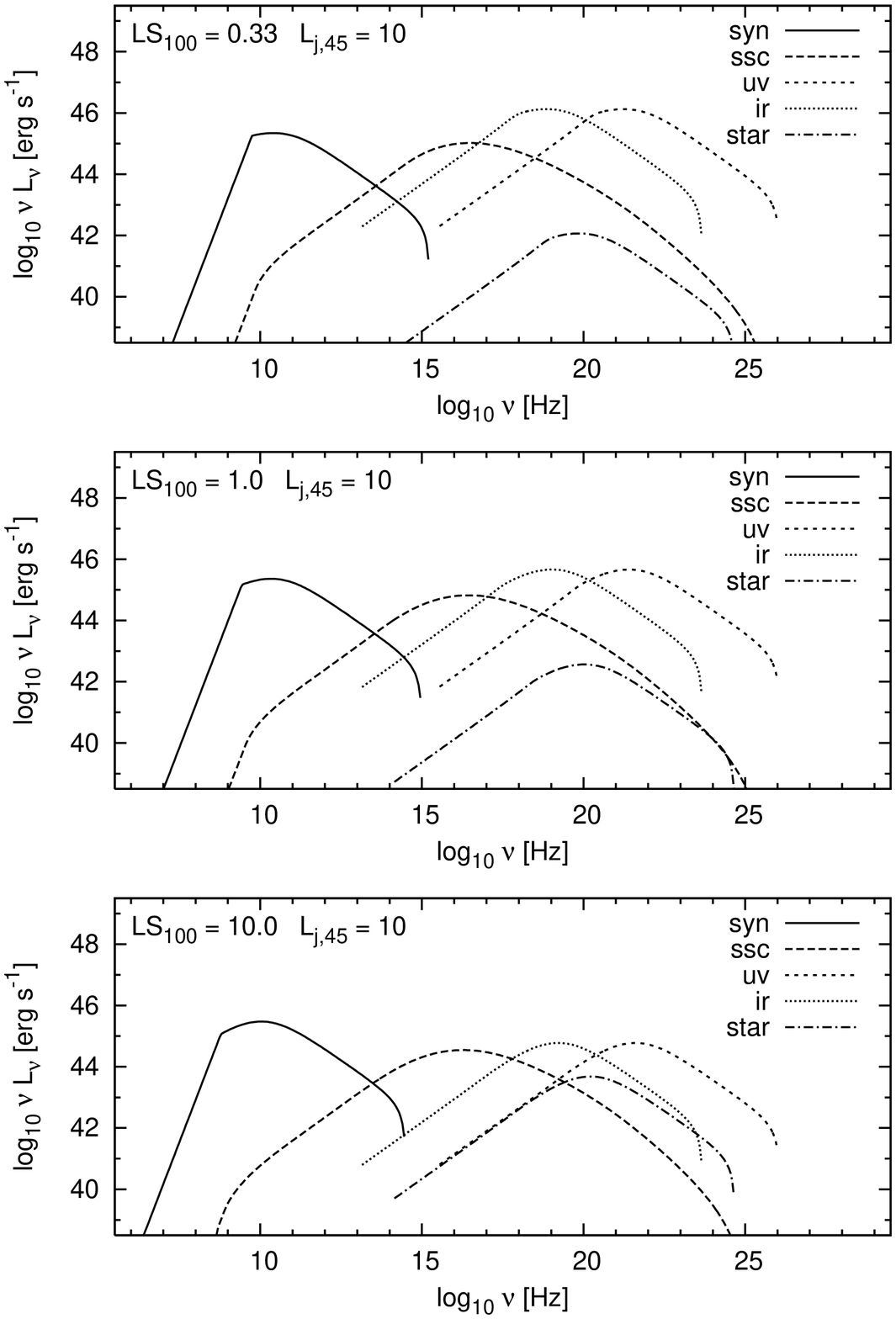}
\includegraphics[scale=0.4]{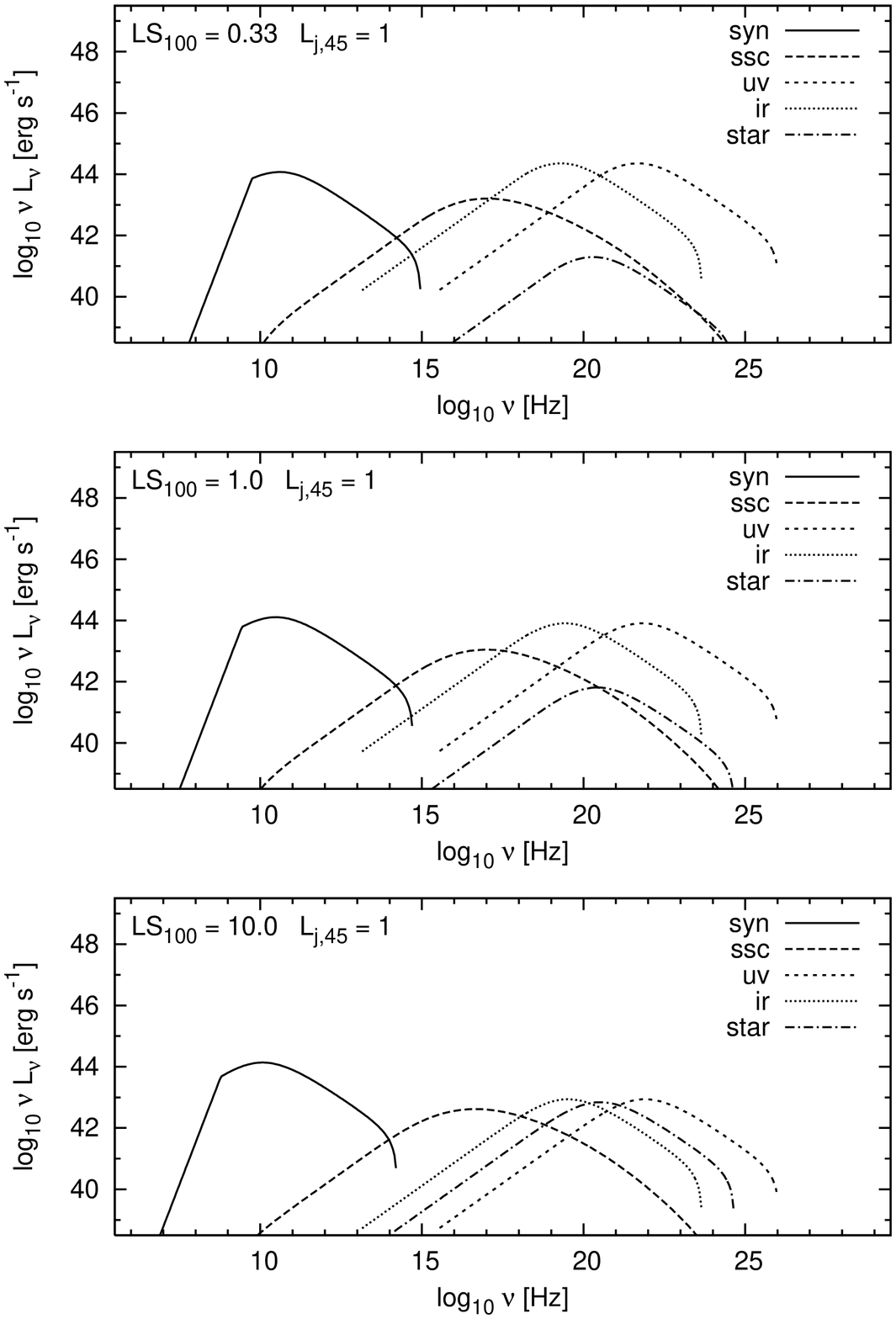}
\includegraphics[scale=0.4]{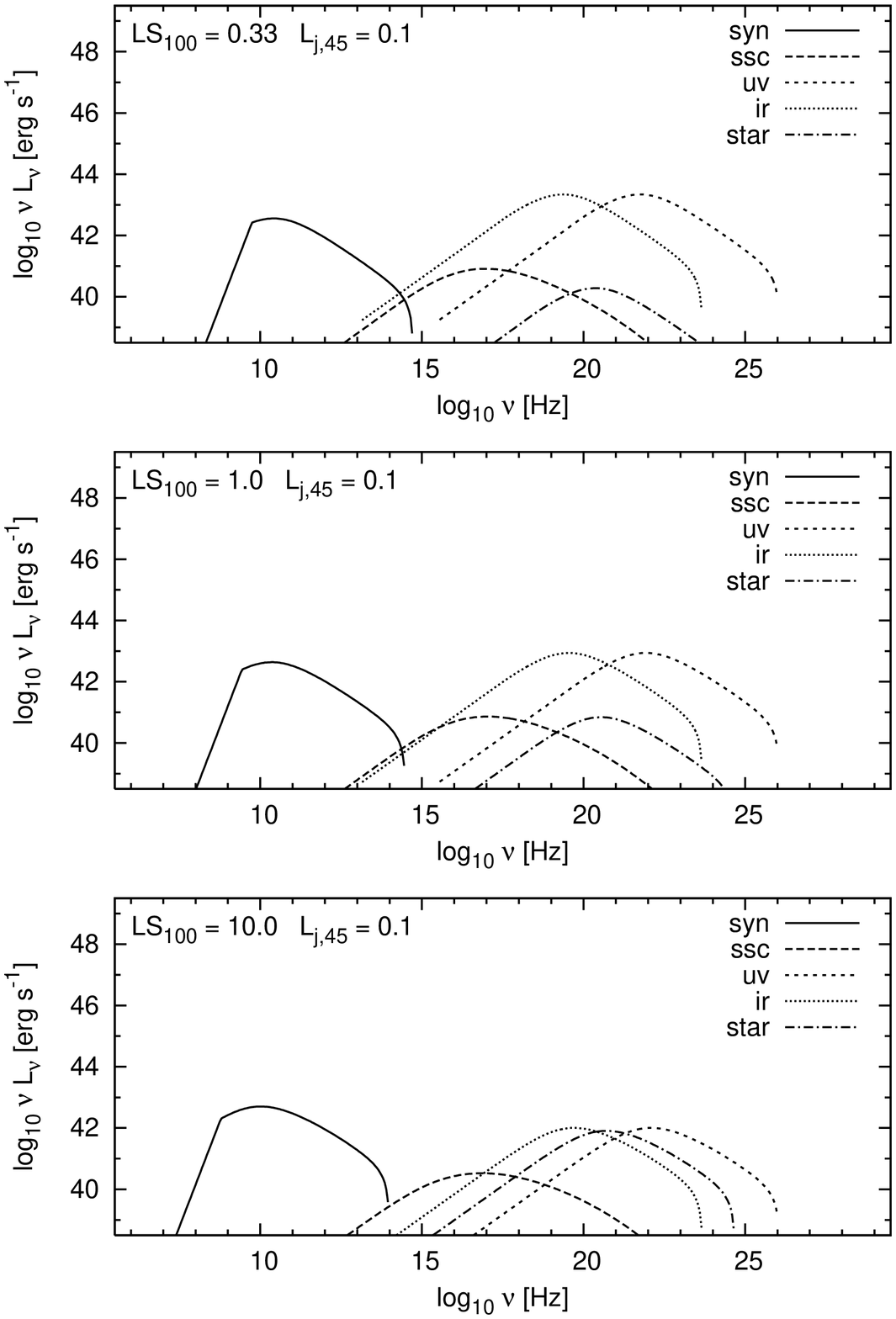}
}
\figcaption{
Broad-band emission produced within the lobes of GPS sources with different jet kinetic power ($L_{\rm j} = 10^{47}$\,erg\,s$^{-1}$, $10^{46}$\,erg\,s$^{-1}$, $10^{45}$\,erg\,s$^{-1}$, $10^{44}$\,erg\,s$^{-1}$) and different linear sizes ($LS=33$\,pc, $100$\,pc, and $1$\,kpc). Illustrative parameters were considered: $\eta_{\rm B} = 0.3$, $\eta_{\rm e} = 3$, $L_{\rm V} = 10^{45}$\,erg\,s$^{-1}$, $L_{\rm UV} = L_{\rm IR} = 10^{46}$\,erg\,s$^{-1}$ for $L_{\rm j} > 10^{45}$\,erg\,s$^{-1}$, and $L_{\rm UV} = L_{\rm IR} = 10^{45}$\,erg\,s$^{-1}$ for $L_{\rm j} \leq 10^{45}$\,erg\,s$^{-1}$. Broken-power-law injection function $Q(\gamma)$ with $s_1 = 1.5$, $s_2 = 3$, break energy $\gamma_{\rm int} = m_{\rm p}/m_{\rm e}$, $\gamma_{\rm min} = 1$, and $\gamma_{\rm max} = 10^5$ was assumed.
} 
\end{figure*}

\end{document}